 \documentclass[final,5p,times,twocolumn,superscriptaddress,secnumarabic,amssymb,amsmath,nobibnotes,nofootinbib]{elsarticle}
 
\usepackage[T1]{fontenc}
\usepackage[utf8]{inputenc}
\usepackage{graphicx}
\usepackage{hyperref}
\usepackage[toc]{appendix}
\usepackage{multirow}
\usepackage{amsmath}
\usepackage{amsthm}
\usepackage{amsfonts}
\usepackage{xcolor}
\usepackage{academicons}
\usepackage{orcidlink}
\usepackage{amssymb}
\usepackage{stmaryrd}
\usepackage{soul}
\usepackage{cleveref}
\usepackage{mathtools}
\usepackage[shortlabels]{enumitem}

\newtheorem{theorem}{Theorem}[section]

\newcommand{\orcid}[1]{\href{https://orcid.org/#1}{\textcolor[HTML]{A6CE39}{\aiOrcid}}}
\newcommand{\abs}[1]{\vert{#1}\vert}

\DeclareMathOperator{\diag}{diag}

\def\be{\begin{equation}}
	\def\ee{\end{equation}}
\def\bea{\begin{eqnarray}}
	\def\eea{\end{eqnarray}}

\newtheorem{proposition}[theorem]{Proposition}
\newtheorem{example}[theorem]{Example}
\newtheorem{remark}[theorem]{Remark}

\usepackage{lipsum}

\journal{Physics Letters B}

\begin{document}
\def\ra{\longrightarrow}
	\def\ci{\mathcal I}
	\def\ca{{\mathcal A}}
	\def\cb{{\mathcal B}}
	\def\cc{{\mathcal C}}
	\def\cd{{\mathcal D}}
	\def\ch{{\mathcal H}}%
	\def\cm{{\mathcal M}}
	\def\cv{{\mathcal V}}
	\def\cw{{\mathcal W}}
	\def\cR{{\mathcal R}}
	\def\hr{{\hat R}}
	\def\cH{{\mathcal H}}
	\def\fH{{\mathfrak H}}
	\def\cT{{\mathcal T}}
	\def\da{{\dot{a}}}
	\def\dph{{\dot{\Phi}}}
	\def\aa{{\mathfrak a}}
	\def\xx{{\mathfrak y}}
	\def\tT{{\mathfrak t}}
\begin{frontmatter}



\title{Is it possible to separate baryonic from dark matter within the $\Lambda$-CDM formalism?}

\author[first]{Andrzej Borowiec \orcidlink{0000-0002-8146-1188}}
\ead{andrzej.borowiec@uwr.edu.pl}
\author[second]{Marcin Postolak \orcidlink{0000-0003-4868-6358}}
\ead{marcin.postolak@uwr.edu.pl}
\affiliation{University of Wroclaw, Institute of Theoretical Physics, pl. Maxa Borna 9, 50-206 Wroclaw, Poland}

\begin{abstract}
We found general solutions of matter stress-energy (non-)conservation in scalar-tensor FLRW-type cosmological models by extending the logotropic formalism to the case of non-minimal coupling between the scalar field and new dark fluid candidates. The energy conditions expressed by the generating function are introduced. Next, we investigate the possibility of separating baryonic from dark matter and explain their ratio as a chameleon effect in the presence of non-minimal coupling. To answer the question affirmatively we analyze simple extensions of $\Lambda$-CDM model by adding a non-minimally coupled scalar field in the Einstein frame. Two scenarios involving either a scalaron (quintessence) or a phantom (ghost) are numerically solved and compared. As a result, it is shown that in both cases LCDM model can be reproduced with a high accuracy in the region covered by observations. We have also demonstrated the compatibility of the two models under consideration with available PPN parameters estimations. As expected, in the case of the phantom (ghost) field the Big-Bang scenario is replaced by the (matter) Bounce.
\end{abstract}




\end{frontmatter}



\section{Introduction}
Modern relativistic cosmology is one of the most rapidly evolving branches of physics and science in general. The discovery of the late-time accelerating expansion of the Universe \cite{Riess1998,Perlmutter1999} is one of the most fundamental challenges facing modern theoretical physics. These questions are partially addressed by the $\Lambda$-CDM (LCDM) model \cite{Frieman2008}\footnote{The reader will find more details in the textbooks \cite{PeeblesPrinciples,Mukhanov2005}.}\footnote{A systematic and chronological overview of the evolution of physical cosmology throughout the 20th and early 21st centuries can be found in \cite{PeeblesCentury}.}. The next step was the proposal of an inflation mechanism\footnote{For a more extensive discussion of various inflationary scenarios, see \cite{Brandenberger1999}.} \cite{Guth1981,Starobinsky1980,Martin2014} designed to explain the flatness and horizon problem. This mechanism, together with the reference model, constitute a kind of paradigm, largely consistent with observational data, such as those involving CMB analysis associated with the \textit{Planck mission} \cite{Planck2018parameters,Planck2018CMB,Planck2018inflation} and measurements of type Ia supernovae in \textit{Pantheon+} \cite{Pantheon+data,Pantheon+constraints}. However, it is worth noting that this approach also has its weaknesses, such as the fine tuning, the unclarified nature of the inflaton, and the lack of a unified and consistent description of the dark sector of the Universe - dark matter \cite{Bertone2010,Bertone2018RMP,Bertone2018Nature,Arbey2021} and dark energy \cite{Bamba2012}\footnote{For an overview of the problems posed to the LCDM model, see \cite{Perivolaropoulos2022}.}.

The current description of dark matter is plagued by several important inaccuracies in galactic scales, such as: cuspy halo problem, dwarf galaxy problem, satellite disk problem, galaxy morphology problem and one of the most relevant issues from the cosmological point of view - high redshift galaxies (such as \textit{JADES-GS-z13-0} and \textit{JADES-GS-z14-0} observed by the James Webb Space Telescope at $z=13.20^{+0.04}_{-0.07}$ \cite{Curtis2023} and $z=14.32^{+0.08}_{-0.20}$ \cite{Carniani2024} respectively).

Despite the obvious advantages and triumphs associated with the combination of the $\Lambda$-CDM model and the inflationary mechanism, one of the most rapidly growing branches of physics is an approach that searches for modifications and deviations from the classical general theory of relativity and cosmological inflation \cite{Brandenberger2020}. Scientists are making every effort to improve the paradigm, or to propose new hypotheses describing our Universe (for current papers on this topic, see \cite{Bull2016,Abbott2023,Aluri}). Note, also, that the motivation regarding the introduction of brand new theories in general also extends to much more fundamental questions about gravity and cosmology itself\footnote{A discussion of these issues can be found in \cite{Shankaranarayanan2022}.}. The issue of modified gravity has received a significant amount of attention in many review papers \cite{Capozziello2008,Capozziello2011,Nojiri2017,Clifton2012,Heisenberg2019,Saridakis2021,Odintsov2023Inflation}, as well as many tests of the compatibility of these proposals with observational data have been carried out (e.g. \cite{Koyama2016})\footnote{A collective description of many aspects of modified gravity is provided in \cite{Kazuya2020}.}.

One of the most widespread as well as recognized attempts to modify GR are the $f(R)$ theories \cite{deFelice,Chen2009,Harko2018,Bertolami2007}, which replace the standard term in the action associated with the curvature scalar by a function depending on it. Meanwhile, these theories can be related to another class of modified gravity theories - scalar-tensor theories (STT) \cite{Fujii2003,Faraoni2004,Quiros2019}. In certain situations, these approaches are equivalent to each other, however, it is not always possible to link these two formalisms (for more details, see \cite{Faraoni2004,fR_STT}). However, even within STT, we are dealing with so-called conformal frames, since the scalar field that is an additional degree of freedom in our theory can be non-minimally coupled to the gravitational segment (Jordan frame) or the matter part (Einstein frame) within the action of the theory. There is no clear consensus on which of these frames is physical. Opinions on this question are sharply divided among cosmologists \cite{Dicke1962,Faraoni1998,Faraoni1999,Flanagan2004,Faraoni2007,Capozziello2010,Chiba2013,Salvio2021}.

Moreover, what is worth emphasizing in the case of the chameleon mechanism (SF non-minimally coupled to the matter in the Einstein frame), it is possible to verify it. Extensive experimental tests are being conducted on the detection associated with \cite{Burrage2018}: 5th force (torsion balance, Casimir effect), atom interferometry, neutrons (neutron interferometry, ultra-cold neutrons), astrophysics (distance indicator, rotation-curve, galaxy clusters), $f(R)$ gravity (CMB, Solar System bounds) and coupling to photons (ADMX, CAST). The results of these studies may shed new light on the problem of conformal frames in cosmology.

Another popular approach in recent years is the attempt to unify dark matter and dark energy within the so-called \textit{dark fluid}. Under this formalism, the dark sector of the Universe is assumed to be a single physical phenomenon. At galactic scales it reproduces the behavior of dark matter, while at cosmological scales it reconstructs the evolution of dark energy. In many cases, equations of state motivated by specific examples of the wide range of solid state physics provide a good basis for formulating dark fluid equations of state\footnote{The reader can find more specific examples in Proposition~\ref{AB}.}, e.g. recent model with cosmological fluid reproducing \textit{Murnaghan} EoS \cite{Murnaghan1944}\footnote{The Murnaghan EoS models the behavior of matter under conditions of high pressure and states that at $T=\mathrm{const}$ the bulk modulus of the incompressibility $K=-V{\left(\frac{\partial p}{\partial V}\right)}_{T}$ is a linear function of pressure:\\
$p=\frac{K_{0}}{K'_{0}}\left[\left(\frac{V}{V_{0}}\right)^{-K'_{0}}-1\right];\quad K'\equiv\frac{\partial K}{\partial p}$.} of the following form \cite{Dunsby2024}:
\begin{equation}
    p=-\frac{A_{*}}{\alpha}\left[\left(\frac{\rho_{*}}{\rho}\right)^{\alpha}-1\right]\sim \rho^{-\alpha}
\end{equation}
corresponding to Chaplygin-like behavior. More details about the motivation and mathematical structure of the above formalism can be found in \cite{Dunsby2016,Luongo2018,Dagostino2022}.

The purpose of our paper is twofold:
\begin{enumerate}
    \item Study of stress-energy non-conservation in ST FLRW cosmological model in the context of the so-called chameleon mechanism, providing a general solution to this problem.
    \item Proposing toy models: relatively simple modifications of the LCDM by adding a non-minimally matter-coupled scalar field as an object that effectively describes the dark matter phenomenon and is able to explain the dark-to-baryonic matter ratio. The segment related to dark energy (cosmological constant) remains unaffected, so this article does not aim to explain this feature.
\end{enumerate}

In Section~\ref{Scalar-tensor gravity}, following \cite{Borowiec2020,Borowiec2021} we recall the formalism associated with ST cosmology in the most generic case including non-minimal coupling (NMC) between gravity and matter as a realization of the   chameleon mechanism \cite{Veltman1977,Khoury20042} and a corresponding stress-energy non-conservation. Furthermore, we propose general solution to a comoving fluid non-conservation in the FLRW background in terms of arbitrary generating function  describing energy density and pressure. In addition, we reformulate the standard energy conditions known from relativistic cosmology in terms of generating function.

In Section~\ref{Toy models}, we propose two toy cosmological models: the first corresponding to the minimal extension of the LCDM model with \textit{scalaron} field (with the presence of an initial singularity) and the second being a concrete realization of the alternative to cosmic inflation \cite{Postolak:2024xtm} - \textit{matter bounce} scenario. Based on the numerical solutions, we analyze the time evolution of two models and try to gain the possible physical outcomes. The analysis of PPN parameters related to the constraints placed on modified theories of gravity was also carried out.

\subsection{Scalar-tensor gravity \& FLRW cosmology}\label{Scalar-tensor gravity}

Our starting point is the most general action for scalar-tensor theories of gravity  which can be defined as follows (see e.g., \cite{Borowiec2020,Borowiec2021,Jarv} for the same convention):
\begin{equation}\label{general action}
	\begin{split}
		& S[g_{\mu\nu},\Phi,\chi] = \frac{1}{2\kappa^2}\int d^4x\sqrt{-g} \Big[\ca(\Phi)R -\cb(\Phi)g^{\mu\nu} \\
		& \partial_\mu\Phi\partial_\nu\Phi-\mathcal{V}(\Phi)\Big] + S_\mathrm{matter}\left[e^{2\alpha(\Phi)}g_{\mu\nu}, \chi\right],
	\end{split}
\end{equation}
where:
$\{\ca(\Phi), \cb(\Phi), \cv(\Phi), \alpha(\Phi)\}$ are the four arbitrary functions so-called frame parameters.\\
Usually $ \ca(\Phi) $ is a positive function of the scalar field $ \Phi $, which is referred to as the so-called effective gravitational constant and which is referred to as the non-minimal coupling of $\Phi$ to gravity. The $\cb(\Phi)$ function describes a non-canonical kinetic term associated with the scalar field  and $\cv(\Phi)$ is the self-interaction potential of the scalar field itself. $\alpha(\Phi)\ $ in turn, is responsible for yet another non-minimal coupling of the scalar field $\Phi$ to the matter fields $\chi$. All of them appear naturally when one passes from $f(R)$ gravity into its scalar-tensor representation. Particularly, in the Einstein frame one finds that $\alpha'(\Phi)\neq 0$ (e.g. \cite{Borowiec2020}).

By performing a variation of the action \eqref{general action} with respect to the metric tensor $ g_{\mu\nu} $ and the scalar field $ \Phi $, we obtain the field equations describing our theory: 
\begin{subequations}\label{STT general eq}
	\begin{align}
		\begin{split}
			& \ca G_{\mu\nu} + \left(\frac{1}{2}\cb  + \ca''\right)g_{\mu\nu}g^{\alpha\beta}\partial_\alpha\Phi\partial_\beta\Phi -\left(\cb + \ca''\right)\partial_\mu\Phi\partial_\nu\Phi \\
			& +\ca'(g_{\mu\nu}\Box - \nabla_\mu\nabla_\nu)\Phi +\frac{1}{2}\cv g_{\mu\nu} = \kappa^2 T_{\mu\nu},
		\end{split} \\
		\begin{split}
			& 2\Big[3(\ca')^2 +2\ca\cb\Big]\Box\Phi + \Big[2\Big(\ca\cb'+\ca'(\cb+3\ca'')\Big)\Big](\partial\Phi)^2\\
			& +2(2\ca'\cv-\ca\cv')=2\kappa^2 T (\ca'-2\alpha' \ca),
		\end{split}
	\end{align}
\end{subequations}
where:
$ ()'\equiv\dfrac{d}{d\Phi} $, $ \Box=g^{\mu\nu}\nabla_\mu \nabla_\nu $ and $T_{\mu\nu}=-\frac{2}{\sqrt{-g}}\frac{\delta S_m}{\delta g^{\mu\nu}}$ denotes stress-energy tensor representing an external matter source with $T=g^{\mu\nu} T_{\mu\nu}$.

Some relevant observations can be deduced from relation \eqref{STT general eq}. First, when $ 3(\ca')^2 +2\ca\cb<0 $ then the scalar field is a ghost\footnote{See Section~\ref{Phantom intro}}. Second, in the case $ \ca'-2\alpha'\ca=0 $ our scalar field $\Phi$ is minimally coupled to matter otherwise it is generated by matter itself. In addition, the condition for the matter stress-energy to be covariantly conserved is the vanishing of derivative from the $ \alpha(\Phi) $ function, i.e. $ \alpha'(\Phi)=0 $.  
In general, from the field equations, it follows that the matter stress-energy tensor is not conserved unless $\alpha'(\phi)=0$ (see, \cite{Borowiec2020,Jarv} for details):
\begin{equation}\label{NMC  continuity}
	\nabla_\mu T^{\mu\nu}=\alpha'(\Phi)\,\partial^{\nu}\Phi\,T\,.
\end{equation}
This is a manifestation of the so-called Chameleon mechanism as introduced in \cite{Khoury20041}.
If this derivative takes nonzero values then the matter particles follow the geodesics of conformally transformed metric:
\begin{equation}
    \tilde{g}_{\mu\nu}=e^{2\alpha(\Phi)}g_{\mu\nu}
\end{equation}
and this generates deviations from the geodesics of $ g_{\mu\nu} $ due to the presence of the so-called "fifth force" associated with the existence of $\Phi$ \cite{Faraoni2004}.

For cosmological applications, one takes the FLRW metric with the scale factor $a(t)$:
\begin{equation}\label{FLRW metric}
	g_{\mu\nu}=\diag\left(-N(t)^2,\frac{a(t)^2}{1-kr^2},a(t)^2 r^2,a(t)^2 r^2 \sin^2{\theta}\right),
\end{equation}
where: $ k=\left\{\pm1;0\right\} $ denotes spatial curvature of the Universe. The laps function $N(t)$ allows control of the reparametrization of the time variable and plays an important role in minisuperspace formulation. The case $N=1$ determines the physical cosmic time.

An external matter source is assumed in a perfect fluid form:
\begin{equation}\label{e-m tensor}
	T_{\mu\nu}=(p+\rho)u_{\mu}u_{\nu}+p g_{\mu\nu},
\end{equation}
where:
\begin{equation}
    u^{\mu}=\left(N^{-1};0;0;0\right)
\end{equation}
represents comoving 4-velocity with conventional normalization $u^\mu u_\mu=-1$.\\
If $\alpha'(\Phi)=0$ then the conservation principle of the matter stress-energy tensor \eqref{NMC  continuity} satisfies  the usual expression:
\begin{equation}\label{MC continuity}
\nabla_\mu T^{\mu\nu}=0,
\end{equation}
that in terms of time-dependent energy density $\rho(t)$ and the pressure $p(t)$ takes very well-known form:
\begin{equation}\label{MC cosmo}
    \dot{\rho}+3H\left(p+\rho\right)=0.
\end{equation}
Here: $\dot{( )}\equiv\frac{d}{dt}$ denotes the differentiation w.r.t. any local coordinate time and $H=\dot{a}/a$ is the Hubble parameter as measured by a local observer. 
It means that the time dependence appears only implicitly as a consequence of the time-rescaling invariance: $dt\mapsto N(t)dt=d\tilde t$. Consequently, solutions ${\rho=\rho(a), p=p(a)}$ are functions of the scale factor. Moreover, the energy density $\rho(a)$ determines the pressure:
\begin{equation}
    p(a)=-\frac{a\,d\rho(a)}{3da}-\rho(a).
\end{equation}
Similarly, in the presence of non-minimal coupling ($\alpha'(\Phi)\neq 0$)  the matter  stress-energy non-conservation  \eqref{NMC  continuity} takes in the FLRW background the following time reparametrization invariant form:
\begin{equation}\label{NMC cosmo}
    \dot{\rho}+3H\left(p+\rho\right)=-\dot{\alpha}\left(3p-\rho\right)
\end{equation}
and, moreover, it is expected to have solutions as explicit functions ${\rho(a,\Phi), p(a,\Phi)}$ (see Proposition~\ref{AB}).

Now substituting back to the field equations \eqref{STT general eq} and
assuming a spatially flat Universe with $N=1$ we obtain a closed (over-determined) system of the second-order ODE for two functions $a(t)$ and $\Phi(t)$:
\begin{subequations}\label{eq STT our case}
	\begin{align}
		\begin{split}
			& 3H^2  = \frac{\kappa^2\,\rho(a,\Phi)}{\mathcal{A}(\Phi)}  + \frac{\mathcal{B}(\Phi)}{2\,\mathcal{A}(\Phi)}\dot{\Phi}^2 - 3\frac{\mathcal{A}'(\Phi)}{\mathcal{A}(\Phi)}H\dot{\Phi}+
	  \frac{\mathcal{V}(\Phi)}{2\mathcal{A}(\Phi)}\,, \label{ee1}
		\end{split}\\
		\begin{split}
			& 2\dot{H} + 3H^2  = -  \frac{\kappa^2\,p(a,\Phi)}{\mathcal{A}(\Phi)}
			-\frac{\mathcal{B}(\Phi) + 2\mathcal{A}''(\Phi)}{2\,\mathcal{A}(\Phi)} \dot{\Phi}^2\\   
			&\quad\qquad\qquad + \frac{\mathcal{V}(\Phi)}{2\,\mathcal{A}(\Phi)} -\frac{\mathcal{A}'(\Phi)}{\mathcal{A}(\Phi)}\left(2H\dot\Phi + \ddot{\Phi} \right)\, ,\label{ee2}
		\end{split}\\
		\begin{split}
			& \left(3(\mathcal{A}'(\Phi))^2 + 2\mathcal{A}(\Phi)\mathcal{B}(\Phi)\right)\ddot{\Phi}  = -3(3(\mathcal{A}'(\Phi))^2 +\\&
			+ 2\mathcal{A}(\Phi)\mathcal{B}(\Phi))H\dot{\Phi}
   -\left((\mathcal{A}(\Phi)\mathcal{B}(\Phi))' + 3\mathcal{A}'(\Phi)\mathcal{A}''(\Phi)\right)\dot{\Phi}^2
			\\	&\quad\quad +
			 \left(2\mathcal{V}(\Phi)\mathcal{A}'(\Phi) - \mathcal{V}'(\Phi)\mathcal{A}(\Phi)\right) \\
			& \quad\quad + \kappa^2(\rho -3p)(a,\Phi)\left[\mathcal{A}'(\Phi) - 2\alpha'(\Phi)\mathcal{A}(\Phi)\right]\,. \label{ee3}
		\end{split}
	\end{align}
\end{subequations}
This system can be then solved and compared with the LCDM model at least numerically, after choosing required frame functions $\{\ca(\Phi),\cb(\Phi), \cv(\Phi), \alpha(\Phi)\}$ and imposing initial conditions that respect the zero Hamiltonian energy constraints \eqref{ee1}.

More exactly, the system \eqref{ee1}-\eqref{ee3} can be equivalently recast into the form of a constrained conservative two-dimensional Hamiltonian system of classical mechanics called minisuperspace (MSS) formalism. 
Cosmological "initial conditions" (Cauchy data) are, in fact, related to the present-day values of cosmological parameters. Normalizing the scale factor ($a_0=1, \dot{a}_0=H_0$) and assuming that the scalar field has no observable dynamics today ($\dot{\Phi}_0=0$) we obtain \cite{Borowiec2021}:
\begin{equation}\label{InitCond}
	3H_0^2  = \frac{\kappa^2\,\rho(a_0,\Phi_0)}{\mathcal{A}(\Phi_0)}  + 
	\frac{\mathcal{V}(\Phi_0)}{2\mathcal{A}(\Phi_0)}.
\end{equation}
The algebraic relation between $\Phi_0$ and $H_0$ does not depend on the kinetic term $B(\Phi)$. Its solutions provide some cosmological scenarios which can be realized in the form of numerical solutions $(a(t), \Phi(t))$.  This is enough to compare numerically with the well-known LCDM scenario and calculate the baryonic to dark matter ratio.

\section{Engineering of stress-energy (non-)conservation in FLRW models}\label{Engineering of stress-energy}
   	 In fact,  the continuity  equation \eqref{MC continuity} as well as discontinuity  one \eqref{NMC continuity}  can be solved in terms of a priori arbitrary differentiable function $f(x)$ which for physical reasons we can assume, e.g., to satisfy the positive energy density condition\footnote{Some energy conditions allow negative energy densities, see Proposition~\ref{Proposition ECs}.}:
  \begin{equation}
      f(x)\geq 0\qquad \mathrm{for}\qquad x\geq 0.
  \end{equation}
  These generate a huge class of potentially \textit{unphysical} solutions whose usefulness can be controlled by additional parameters e.g. energy conditions\footnote{More comprehensive context regarding energy conditions in the case of relativistic cosmology can be found in \cite{VISSER2000,Cattoen2005,Curiel2017}.} or speed of sound. On the other hand, treating STT as an effective description one may expect that some of them are \textit{physically reasonable} from the point of view of yet unknown more fundamental theory or/and additional fields. Now, by direct calculation, one can verify the validity of the following Proposition:
\begin{proposition}\label{AB}
\begin{enumerate}[(I)]
\item Let $f(x)$ be some differentiable function, which will be called a generating function. We set:
\begin{equation}\label{gx}
    g_{}(x)=x f'(x)-f(x),
\end{equation}
where $'\equiv{d\over dx}$. Then the pair:
\begin{equation}\label{p1}
    \begin{dcases}
        \frac{\rho(a)}{\rho_0}=f(a^{-3})\\
        \frac{p(a)}{\rho_0}=g_{}(a^{-3})
    \end{dcases}
\end{equation}
is a solution of \eqref{MC cosmo} for any dimensionfull constants $\rho_0$.\footnote{We assume the speed of light $c=1$. Here, the constant $\rho_0$ is for dimensional reasons and will be further omitted.}
\item Conversely, if $g(x)$ is given, then the generating function can be reconstructed by:
\begin{equation}\label{p2}
    f(x)=x\left(C_{1}+\int \frac{g(x)}{x^2}dx\right)
\end{equation}
where: $C_{1}$ stands for an integration constant, such that the pair:
\begin{equation} \label{p3}  
        \rho(a)=f_{}(x)|_{x=a^{-3}},
        p(a)=g_{}(x)|_{x=a^{-3}}
\end{equation}
is a solution of \eqref{MC cosmo}. Moreover, both transformations are inverse to each other.
\item If functions $\rho=\rho(a)$ and $p=p(a)$ are solutions of \eqref{MC cosmo} then:
\begin{equation}\label{p4}
    \begin{dcases}
        \rho(a)\mapsto \rho_{\mathrm{}}(a,\Phi)=e^{4\alpha(\Phi)}\rho\left(a\,e^{\alpha(\Phi)}\right)\\
        p(a)\mapsto p_{\mathrm{}}(a,\Phi)=e^{4\alpha(\Phi)}\,p\left(ae^{\alpha(\Phi)}\right)
    \end{dcases}
\end{equation}
are solutions of \eqref{NMC cosmo}.
\end{enumerate}
\end{proposition}
Particularly, within ST gravity, it is always possible to find out a conformally related frame in which the stress-energy tensor is conserved, e.g. Jordan frame in $f(R)$ gravity models.	

The formulae relating function $f(x)$ and $g(x)$ are linear as expected, therefore allowing for the creation of composite multi-component objects as a linear combinations of given ones.   
Its first part generalizes the results of \cite{Chavanis2016}. The second one \eqref{p4} is 
to the best of our knowledge new. For the special case of barotropic fluids it has been explored in \cite{Borowiec2021}. It allows to study chameleon mechanism \cite{Veltman1977} as an effect of non-minimal coupling between gravity and matter.

The generalization to the FLRW spacetime in arbitrary Lorentzian dimension $n+1$ is also possible:
\begin{equation}\label{new_n}
	\begin{dcases}
		\rho(a)\mapsto \rho_{\mathrm{}}(a,\Phi)=e^{(n+1)\alpha(\Phi)}\rho\left(a e^{\alpha(\Phi)}\right)\\
		p(a)\mapsto p_{\mathrm{}}(a,\Phi)=e^{(n+1)\alpha(\Phi)}\,p\left(a e^{\alpha(\Phi)}\right),
	\end{dcases}
	\end{equation}
where:
\begin{equation}\label{new_n2}	 
		\rho(a)=f_{}(x)|_{x=a^{-n}}\,,\quad
		p(a)=g_{}(x)|_{x=a^{-n}}\,.	 
\end{equation}
The formula \eqref{new_n} is a solution of the  stress-energy non-conservation:
\begin{equation}
	\dot{\rho}+n\,H\left(p+\rho\right)=-\dot{\alpha}\left( n\,p-\rho\right)
\end{equation}
provided \eqref{new_n2} satisfies the case $\dot{\alpha}=0$.\\

It should be remarked that a similar expression to \eqref{gx}  is used 
for a definition of  the self-interacting potential:
\begin{equation}\label{F_R}
	V(R)=R f'(R)-f(R)\,,
\end{equation} 
see e.g. Appendix 1 in \cite{Borowiec2020}, when changing from $f(R)$-gravity to its ST equivalent, where $R$ is a Ricci scalar. The difference is that above $R=R(\Phi)$ is understood as an inverse function to $\Phi=f'(R)$ while in \eqref{gx} we do not use the Legendre transformation, c.f. \cite{Camera2023}.

We finish general consideration by providing few examples.
	 \begin{example}
	 	 In fact, every choice  $f(x)$ generates two parameter extensions $f_{A,B}(x)=f(A+Bx)$ with $g_{A,B}(x)=Bxf'(A+Bx)-f(A+Bx)$, $B\neq 0$. For one of the simplest cases:
        \begin{equation}
            f_{A,B}(x)=(A+Bx)^{(1+\omega)} 
        \end{equation}
        one gets:
        \begin{equation}
            g_{A,B}(x)=(\omega\, Bx-A)(A+Bx)^{\omega},
        \end{equation}
        which for $A=0$ provides a barotropic fluid with the barotropic equation of state (EoS) parameter $\omega$. Furthermore, for $\omega=-1$ one gets a cosmological constant.  The case $\omega=0$ provides dust matter. Thus:
        \begin{equation}\label{wchameleon}
        \rho_{0,1,\omega}(a,\Phi)=\rho_{0 \omega}\,e^{(1-3\omega)\alpha(\Phi)}\,a^{-3(1+\omega)}	\end{equation}
        as already obtained in \cite{Borowiec2021}.\\
        \textit{It shows that the chameleon factor depends, in fact, on a barotropic parameter $\omega$}.
	 \end{example}
 Originally, the chameleon mechanism has been proposed as an effect of non-minimal coupling between gravity and matter within scalar-tensor gravity formalism \cite{Khoury20041,Khoury20042} as a being related with the change from Jordan to Einstein's frame. In some approaches, it is ad-hoc assumed that it manifests itself as a multiplicative factor in front of the matter Lagrangian that depends on the scalar field. As we can see from expression \eqref{wchameleon} such an assumption is valid only in the case of barotropic EoS. 
 In addition, the multiplicative factor $e^{(1-3w)\alpha(\Phi)}$ heavily depends on the barotropic coefficient $\omega$. However, such formula is not longer true for more general fluids with nonlinear relations between $\rho(a,\phi)$ and $p(a,\Phi)$.
	 \begin{example}
	 	More generally, considering:
        \begin{equation}
        	 \begin{split}
            f(x)& =x^{(\omega+1)}\left[A+Bx^{-(\beta+1)(\omega+1)}\right]^{\frac{1}{\beta+1}}\\
           &= \left[B+Ax^{(\beta+1)(\omega+1)}\right]^{\frac{1}{\beta+1}}\,,
        \end{split}
        \end{equation}
      one obtains:
        \begin{equation}
            g(x)=\omega f(x)+(1+\omega)\frac{B}{f(x)^{\beta}},
        \end{equation}
        which provides a generalized Chaplygin gas, see e.g. \cite{Bento2002GCG}. In particular, for $B=0$ one recovers barotropic fluid and for $\beta=1, \omega=0$  the standard Chaplygin gas \cite{Kamenshchik2001CG}.
	 \end{example}	 
\begin{example}
	The choice:
    \begin{equation}
        f(x)=\frac{x^\alpha}{(\alpha-1)^2} (\ln x^{\alpha-1}-1), \qquad \alpha\neq 1
    \end{equation}
    implying
    \begin{equation}
        g(x)=x^{\alpha}\ln{x}
    \end{equation}
    is known a generalized logotropic case \cite{Benaoum2022}.\\
    The Anton-Schmidt fluid, see \cite{Capozziello2018,Capozziello2019AS,Boshkayev2019}, is obtained for
    \begin{equation}
      f(x)=\frac{x}{2}\ln^2 x\,,   \quad g(x)=x\ln x\,.
    \end{equation}
\end{example}
 
\subsection{Energy conditions and other applications}\label{Energy conditions and other applications}
In this subsection, we discuss some constraints on the generating function $f(x)$ that comes from its possible physical interpretation as an energy density.

Firstly, notice that choosing $g(x)$ as a generating function, one gets:
\begin{equation}
    \frac{\rho(x)}{\rho_0}=Dx+Cx\int{\frac{g(A+x)}{x^2}dx}
\end{equation}
allowing, as observed in \cite{Chavanis2016}, to infer the first law of thermodynamics:
\begin{equation}
    xd\rho=(\rho+p)dx=Bxf'(A+Bx)\big|_{x=a^{-3}}\,.
\end{equation}
Further, we can introduce two quantities characterizing the physical properties of a fluid; an effective barotropic factor $w(f)$, also known as the equation of state (EoS) parameter and (effective) speed of sound $c_s(f)$ as functions of the scale factor by the following expressions:
\begin{equation}\label{feos}
	w(f_{})=\frac{p}{\rho}=-1+\frac{x f'(x)}{f(x)}\bigg |_{x=a^{-3}},
\end{equation}
\begin{equation}\label{fsound}
	c_s^2(f_{})=\frac{dp}{d\rho}=\frac{x f''(x)}{f'(x)}\bigg |_{x=a^{-3}}\,.
\end{equation}
Particularly, $c_s^2(f_{})=1$ for $f(x)=x^2$, i.e. for a stiff matter. 
Matter-dominated era $\omega(f)\approx 0$ means that
\begin{equation}
	xf'(x)\approx f(x)
\end{equation}
for the wide range of $x$, i.e. a dust matter.

Further constraints on the generating function $f(x)$ can be imposed by the so-called energy-conditions:
\begin{equation}\label{gec}
    \mathrm{DEC} \subset \mathrm{WEC} \subset \mathrm{NEC} \supset \mathrm{SEC}\,.
\end{equation} 
More exactly, under the reasonable assumption that:
\begin{equation}
    0\leq x=a^{-3}<\infty,
\end{equation}
one gets the following\footnote{Energy conditions can also be understood pointwise, as limiting spacetime regions, which in the context of FLRW cosmology means restricting the time variable.}:
\begin{proposition}\label{Proposition ECs}
Energy conditions in terms of generating function:
\begin{enumerate}
	\item Dominant energy condition (DEC):
    \begin{equation}
        \rho\geq\abs{p}
    \end{equation}
	\begin{equation}\label{dec}
		2f(x)\geq xf'(x)\geq f(x)\geq 0
	\end{equation}
	or
	\begin{equation}\label{dec2}
		f(x)\geq xf'(x)\geq 0.
	\end{equation}
	The first equality $2f(x)= xf'(x)$ holds for a stiff matter $f(x)=Ax^2$ while the second 
	$f(x)= xf'(x)$ holds for a dust $f(x)=Ax$, $A>0$.
	\item
	Weak energy condition (WEC):
    \begin{equation}
        \begin{dcases}
            \rho\geq0\\
            \rho+p\geq0
        \end{dcases}
    \end{equation}
	\begin{equation} \label{wec}
		f(x)\geq0  \wedge f'(x)\geq0.
	\end{equation}
    is satisfied, for instance, if the cosmological constant is positive.
	\item 	Null energy condition (NEC):
    \begin{equation}
        \rho+p\geq0
    \end{equation}
	\begin{equation}\label{nec}
		f'(x)\geq0.
	\end{equation}
	is e.g, satisfied by positive and negative cosmological constants.	 
	\item
	Strong energy condition (SEC):
    \begin{equation}
        \begin{dcases}
            \rho+p\geq 0\\
            \rho+3p\geq 0
        \end{dcases}
    \end{equation}
	\begin{equation}\label{sec}
		f'(x)\geq0\wedge
		f'(x)\geq\frac{2}{3}\frac{f(x)}{x}.
	\end{equation}
	Here, the equality holds for positive spatial curvature (cosmic strings):
    \begin{equation}
        f(x)=A\,x^{2/3}
    \end{equation}
    with $A>0$.
\end{enumerate}
\end{proposition}
It can be noticed that NEC \eqref{nec} and SEC \eqref{sec} admit negative energy densities, and therefore, negative values for the generating function $f(x)$ for $x\geq 0$.
However, it is not entirely clear how this fact could manifest itself in physical reality\footnote{An attempt to answer this question has been included in the paper \cite{Nemiroff2015}.} (perhaps it would be a footprint of higher-dimensional/quantum gravity theories). Also
energy density $\rho(a)$ observed by a co-moving observer can be negative\footnote{General background regarding negative energy densities in terms of quantum effects related to gravity in the case of Hawking radiation can be found in \cite{Penrose2005}.}. 
In STT one should take into account the scalar field dynamics which also contribute to the overall energy balance (see below).

\begin{remark}
Except the barotropic EoS \eqref{wchameleon}, the formulas \eqref{feos} and \eqref{fsound} are changed when non-minimal coupling is active, i.e. when
$x\mapsto a^{-3}e^{-3\alpha(\Phi)}$. 
Similarly, for the energy conditions listed above.
\end{remark}
\begin{remark}
	As follows from Proposition~\ref{AB}, the Friedmann equation in the form:
	\begin{equation}\label{E1}
		3H^2 = \kappa^2 \rho_{f0}\,f(a^{-3})\equiv \kappa^2 \rho_f
	\end{equation}
	where $f(x)$ is, in principle, any differentiable function is consistent with the Einstein equation ($E_{11}$ component):\footnote{The dimensional integration constant $\rho_{f0}$ and the gravitational constant $\kappa^2=8\pi G$ are necessary for dimensional reasons. Further we work with geometric units $\kappa=c=1$.}
	\begin{equation}\label{E2}
		3H^2+2\dot{H} =  -\kappa^2 \rho_{f0}\,\Big(a^{-3} f^\prime(a^{-3})-f(a^{-3}) \Big) \equiv -\kappa^2 p_f
	\end{equation}
through conservation law \eqref{MC cosmo}.\\
Furthermore, assuming:
 \begin{equation}
     f(x)=\sum_{n\geq 0}f_n x^n
 \end{equation}
 to be analytic gives:
 \begin{equation}
     g(x)=-f_0 +\sum_{n\geq 2}(n-1)f_n x^n.
 \end{equation}
 It contains the terms with EoS parameter $w_n=n-1\geq 0$. In this way one can get only late acceleration  provided by the cosmological constant $f_0>0$. Nevertheless, non-analytic functions turn out to be more useful, for example:
 \begin{equation}
     f(x)=\begin{dcases}
         A e^{-B/x};\ \mathrm{for}\ x>0\\
         0;\ \mathrm{otherwise},
     \end{dcases}
 \end{equation}
 with: $A,B>0$.\\
 Moreover, the choice:
	\begin{equation}\label{nonanalytic}
		H^2 = \Lambda + 
		A\,e^{-Ba^3}\,,
	\end{equation}
	enforces an early de Sitter era by $\Lambda\mapsto \Lambda +A$ leaving  the late governed by $\Lambda$.   
\end{remark}

\subsection{Einstein frame action, equations of motion and conservation law}
In the Einstein frame the action takes a very simple form:
\begin{equation}\label{Einstein frame action}
	\begin{split}
		S^\epsilon [g_{\mu\nu},\Phi,\chi]=& \frac{1}{2\kappa^2}\int{d^4 x\sqrt{-g}\left[R-\epsilon g^{\mu\nu}\partial_{\mu}\Phi\partial_{\nu}\Phi-\mathcal{V}(\Phi)\right]}\\
		& +S_{\mathrm{m}}\left[e^{2\alpha(\Phi)}\, g_{\mu\nu},\chi\right]
	\end{split}
\end{equation}

From the point of view of $f(R)$-gravity the parameter $\epsilon=\{\pm1;0\}$ separates three cases: Palatini ($\epsilon=0$), metric  ($\epsilon=1$) and
hybrid ($\epsilon=-1$), see \cite{Borowiec2021}. From the other hand a value of $\epsilon$ changes the character of scalar field itself. For $\epsilon=\gamma=0$ scalar field has no dynamics and undergoes algebraic constraints, $\epsilon=1$ corresponds to scalaron (quintessence) field, see e.g. \cite{Shtanov2021,Shtanov2022,Pogosian2023} while $\epsilon=-1$ is known as a phantom (ghost) field \cite{Sushkov2004,Krause2012,Sbisa,Dabrowski}. In our work, we will refer to the model with $\epsilon=1$ as \textit{Einstein Frame Scalar-Tensor Scalaron} (EFSTS), and the model with $\epsilon=-1$ as \textit{Einstein Frame Scalar-Tensor Ghost} (EFSTG).

For cosmological applications, we rewrite the action \eqref{Einstein frame action} in a more convenient form of two-dimensional conservative mechanical system  known as minisuperspace (MSS) formulation:
\begin{equation}\label{MSS action}
	S^\epsilon_{\mathrm{MSS}}\left[a,\Phi\right]=\int{dt\Big[-6a\dot{a}^2+
		\epsilon 
		a^{3}\dot{\Phi}^{2}-a^{3}\Big(\cv(\Phi)+2\kappa^{2}\rho_f\Big)\Big]},
\end{equation}
where the matter density:
\begin{equation}\label{fdensity}
	\rho_f=\rho_f(a,\Phi)=  \rho_{f0}\,e^{4\alpha(\Phi)}\, f\left(a^{-3}\,e^{-3\alpha(\Phi)}\right)
\end{equation}
is taken in its most general form allowed by an implementation of the  
chameleon mechanism. Corresponding pressure is explicitly expressed as:
\begin{equation}\label{pf}
\frac{p_f}{\rho_{f0}}=  e^{\alpha(\Phi)}\,a^{-3} f^\prime\left(a^{-3}\,e^{-3\alpha(\Phi)}\right)-e^{4\alpha(\Phi)}\, f\left(a^{-3}\,e^{-3\alpha(\Phi)}\right)\,.
\end{equation}
From the relation \eqref{fdensity}, one can easily obtain partial derivatives of the energy density:
 \begin{equation}\label{ab0}
	\partial_a \rho_f =-3a^{-1}\left(\rho_f+p_f\right)\,,\quad   \partial_\Phi \rho_f=\left(\rho_f -3p_f\right)\alpha'(\Phi)\,.
\end{equation}

The zero Hamiltonian energy condition constraining  the system can be recast into the Friedmann equation \eqref{ee1}:
\begin{equation}\label{ab1}
	H_{}^2=\frac{\rho_\Phi}{3} +\frac{\kappa^2\,\rho_f(a,\Phi)}{3},
	 \end{equation}
where the scalar field energy density and pressure are defined as follows:
\begin{equation}\label{ab2}
	\begin{dcases}
		\rho_\Phi=\frac{1}{2}\Big(\epsilon \dot\Phi^2+\cv(\Phi)\Big)\\
		p_\Phi=\frac{1}{2}\Big(\epsilon \dot\Phi^2-\cv(\Phi)\Big)\,.
	\end{dcases}
\end{equation}
By performing a variation of the action \eqref{MSS action} with respect to the scale factor and the scalar field, respectively, we obtain the following explicit form of the system of dynamical equations: 
\begin{subequations}\label{eq12ab}
	\begin{align}
		\begin{split}
			3H^2+2\dot{H} = & -\kappa^2 p_f-p_\Phi\label{eq1ab}
		\end{split}\\
		\begin{split}\label{eq2ab}
			\epsilon\left(\ddot{\Phi}+3H\dot{\Phi}\right) + &\ {1\over 2}\cv^\prime(\Phi)\,=\,-\kappa^2 (\rho_f-3p_f)  \alpha^\prime(\Phi)\, .
		\end{split}
	\end{align}
\end{subequations}
From the equations \eqref{ab1}-\eqref{ab2} it immediately follows that the conservation law \eqref{MC cosmo}for a total or effective energy density:
\begin{equation}\label{ab3}
		\rho_{\mathrm{eff}}=\rho_\Phi+\kappa^2\rho_f\,,\quad p_{\mathrm{eff}}=
		p_\Phi + \kappa^2 p_f\,.
\end{equation}
is always fulfilled. 
The last equation \eqref{eq2ab} instead can be rewritten in the form:
\begin{equation}\label{ab4}
	\dot\rho_{\Phi}+3H(\rho_\Phi+p_\Phi)\,=\,-\kappa^2 \dot{\Phi}\,\partial_\Phi \rho_f \,.
\end{equation}
Thus conservation of $\rho_\Phi$ is equivalent to the conservation of $\rho_f$, which holds  when non-minimal coupling is absent. In this case, according to the Proposition \eqref{AB} $\rho_\Phi$ can be emulated as perfect fluid ($\epsilon\Box\Phi = -{1\over 2}\cv^\prime(\Phi)$):
\begin{equation}\label{ab5}
	\rho_{\Phi} \,=\, \rho_h\equiv h(a^{-3}) \,
\end{equation}
form some function $h(x)$. It means that the time dependence $\rho_\Phi(t)$ for solutions of field equations can be replaced by $h(a^{-3}(t))$.
In fact, an explicit form of the function $\Phi(a)$ can be deduced from a solution of differential equation presented in \cite{Borowiec2021} (see, section 2.4). For the special choice, $\cv(\Phi)=\cv_0=2\Lambda=\mathrm{const}$, one gets:
\begin{equation}\label{ab6}
 \dot\Phi^2 = A\,a^{-6}\,, 
\end{equation}
resulting in:
\begin{equation}\label{ab7}
    \dot{\Phi}=
    \dot{\Phi}_{0}a^{-3},
\end{equation}
where $A=\dot{\Phi}^2_{0}$ is integration constant. Thus in such case, $\rho_\Phi$ effectively consists of gravitational constant and stiff matter.
This observation gives some inside into the nature of scalar field dynamics and its role in ST FLRW-type cosmology.

    \section{Toy models mimicking $\Lambda$-CDM with baryonic and dark matter separated}\label{Toy models}
    In this section we investigate cosmological models taking advantage of the mechanism of non-minimal coupling between the scalar field and matter part of the Universe as a way to distinguish between dark and baryonic matters. Both substances, according to LCDM philosophy, are emulated as cosmic dust. The mathematical formalism introduced so far provides a way to carry out such a procedure.
    \subsection{The models}\label{Einstein frame formalism}
       
    As an illustrative example, we propose to analyze two toy models that, in a sense, minimally extend well-known $\Lambda$-CDM model by adding a scalar field either with positive or negative kinetic energy term and reduce self-interaction potential to a cosmological constant $\Lambda$.  
    Therefore, compared with the previous section we specialize in the potential:
    \begin{equation}\label{m1}
    \cv(\Phi)\equiv	V_{\mathrm{DE}}= 2\Lambda\,.
    \end{equation}
    Also, the matter part containing dust and radiation is taken to be the same as in $\Lambda$-CDM:
    \begin{equation}\label{m2}
\rho=\rho(a,\Phi)=\rho_{\mathrm{R0}}a^{-4}+\rho_{\mathrm{BM0}}\,\Phi^{\gamma}\,a^{-3}\,.
    \end{equation}
      However, our additional modification assumes a non-minimal coupling between the scalar field and the matter that is controlled by the function:
      \begin{equation}
          \alpha(\Phi)=\gamma\ln{\Phi}.
      \end{equation}
      The last term expresses our hypothesis that non-minimal coupling can provide the correct \textit{dark-to-baryonic matter ratio} (radiation term remains unaffected, c.f. \eqref{wchameleon}).
 In this way, we utilize the chameleon mechanism: dust matter is described not by the original FLRW metric $g_{\mu\nu}$ but by a new conformally re-scaled "\textit{dark metric}":
\begin{equation}\label{m3}
	\tilde{g}_{\mu\nu}=\Phi^{2\gamma}g_{\mu\nu},
\end{equation}
while the baryonic matter is related to the original one.

The zero Hamiltonian energy condition:
     \begin{equation}\label{m4}
    	\frac{H^2}{H_{0}^{2}}=\epsilon\frac{\dot{\Phi}_{}^{2}}{6H^2_0}+\Omega_\Lambda+ \Omega_{\mathrm{R0}}a^{-4}_{}+\Omega_{\mathrm{BM0}}\, \Phi_{}^{\gamma}\,a^{-3}_{}
    \end{equation}
    at any instant of time $t$, where dimensionless densities: 
    \begin{equation}\label{m5}
    \Omega_{\Lambda}=\frac{\Lambda}{3H^2_0}\,, \quad	\Omega_{w\,0}=\frac{\kappa^2\rho_{w\,0}}{3H^2_0}
    \end{equation}
    are defined in a standard way, and $H_0$ denotes the current value of the Hubble parameter.\\ 
    Subsequently, equations of motion take the following form: 
    \begin{subequations}\label{eq12m}
       \begin{align}
    		\begin{split}
    			3H^2+2\dot{H} = &\  3\Omega_{\Lambda}-\Omega_{\mathrm{R0}}a^{-4}-\frac{\epsilon}{2}\dot{\Phi}^{2}\label{eq1m}
    		\end{split}\\
    		\begin{split}
    			\epsilon\left(\ddot{\Phi}+3H\dot{\Phi}\right) = &\  -3\gamma\Omega_{\mathrm{BM0}}a^{-3}\Phi^{\gamma-1}\,.
    			\label{eq2m}
    		\end{split}
    	\end{align}
    \end{subequations}
 For $\epsilon=0$ the equation \eqref{eq1m} can be integrated to the Friedmann equation of the $\Lambda$-CDM model\footnote{In fact, $\epsilon=0$ forces $\gamma=0$, c.f. \eqref{eq2m}. In such a case there is no CDM effect in \eqref{m4} due to $\gamma=0$ value.}. More generally, assuming only $\gamma=0$ or $\Omega_{\mathrm{BM0}}=0$, the equation  \eqref{eq2m} admits solution $\dot\Phi=\dot\Phi_0 a^{-3}$, c.f. \eqref{ab6}. In this case \eqref{eq1m} integrates into the Friedmann equation with an additional stiff matter term:
    \begin{equation}
   \frac{H^2}{H_{0}^2}=\epsilon\frac{  \dot\Phi_0^{2}}{6H^2_0}\,a^{-6}+\Omega_\Lambda+ \Omega_{\mathrm{R0}}a^{-4}_{}+\Omega_{\mathrm{BM0}}\,a^{-3}_{}.
   \end{equation}
    It contributes proportionally to $\epsilon$ and the current value of $\dot\Phi_0^2$. In what follows we assume: $\dot\Phi_0=0$, $\gamma\neq 0$ and $\Omega_{\mathrm{BM0}}\neq 0$ ($\dot\Phi_0=\gamma= 0$ reproduces again $\Lambda$-CDM). Thus controlling the term $3\gamma\Omega_{\mathrm{BM0}}\Phi^{\gamma-1}\approx 0$ for $a\approx 1$ we do not interfere in the Friedmann equation for most of the observational data regardless of the value of $\epsilon$\footnote{The problems of the CMB spectrum ($a\approx 10^{-3}$) and structure formation should be discussed separately.}.
    Keeping all these in mind, an example of the numerical solution of the above system of ODE equations will form the basis for further analysis of both toy models.
    
    \subsection{Numerical analysis: large scale}
     Our purpose in this section is to analyze more deeply numerical solutions of the above system of ODE  in order to get more inside into these models
     and test our hypotheses\footnote{All of the diagrams were made in Wolfram Mathematica \cite{Mathematica}.}. Before doing this we want to make clear that \textit{numerical solutions should be treated with some care}. Universe evolution, as described by these models, is a particular trajectory in the phase space of some autonomous (Hamiltonian) dynamical system whose stability is not yet clarified: small deviations in the initial conditions could have a large impact on other stages of the evolution. For this reason, some claims, especially the one concerning an early universe, have rather a speculative and preliminary character, even if they agree with a piece of common knowledge. Future research should use more advanced methods, such as a Markov chain Monte Carlo analysis and cosmological perturbations.

\begin{figure}[h]
	\centering
	\includegraphics[width=1\linewidth]{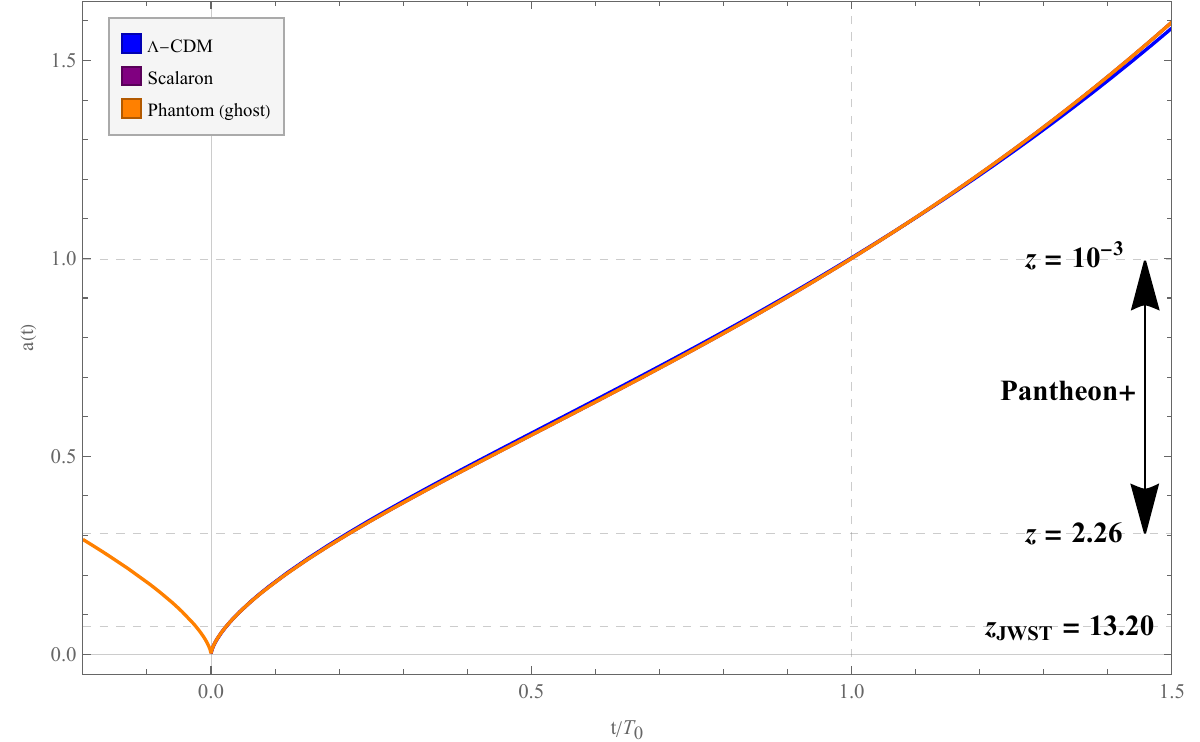}
	\caption{\scriptsize{Evolution of the scale factors for both models reproduces over a large range of time the behavior typical of the LCDM. A cosmic bounce scenario can be observed for EFSTG model ($\epsilon=-1$).}}
	\label{fig:Scale-factor-near-1}
\end{figure}

\begin{figure}[h]
	\centering
	\includegraphics[width=1\linewidth]{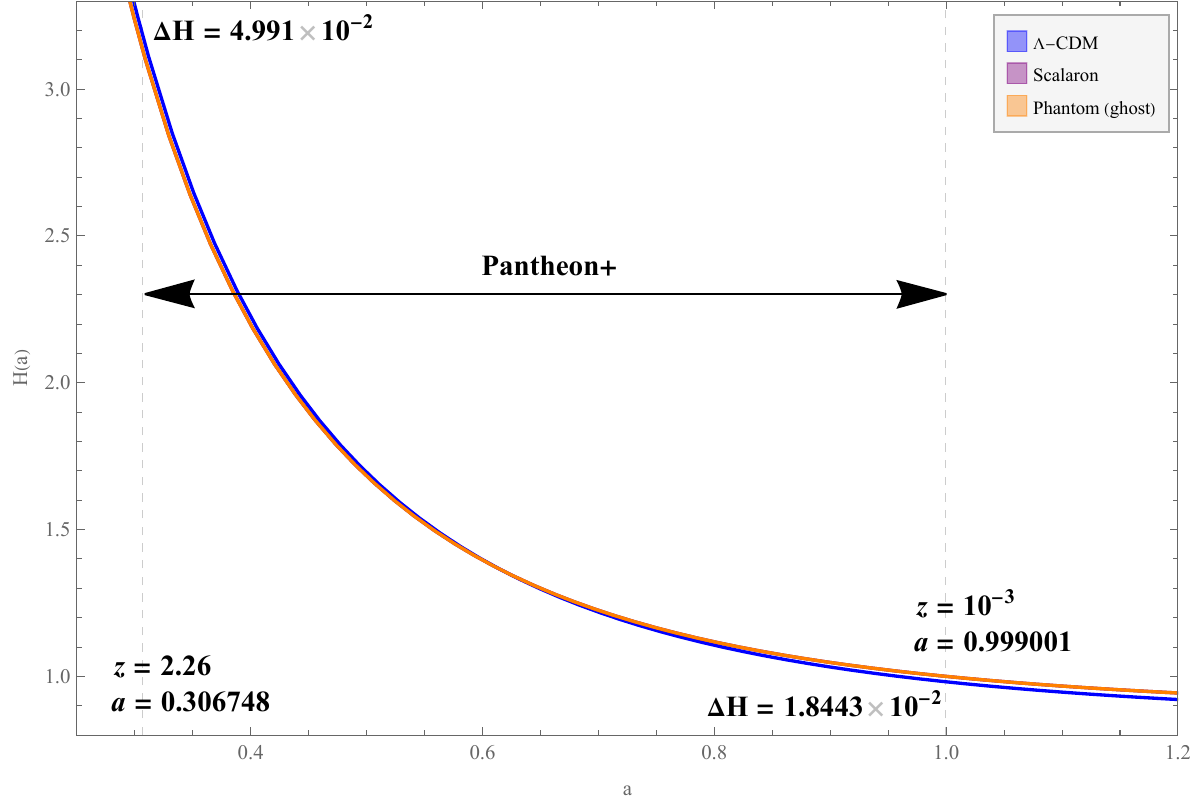}
	\caption{\scriptsize{Evolution of the Hubble parameters for both cosmological scenarios compared with the LCDM model, c.f. \eqref{m4}.}}
	\label{fig:Hubble-parameter-general-difference}
\end{figure}

    First, we notice that the model contains four numerical parameters: $\left\{\Omega_{\mathrm{BM}0}, \Omega_{\mathrm{R}0}, \Omega_\Lambda, \gamma\right\}$ as well as one discreet $\epsilon=\pm 1$ which are explicitly present in the equations of motion \eqref{eq1m}-\eqref{eq2m} we want to solve.
The first two, ie. baryonic matter and radiation densities:
\begin{equation}
    \Omega_{\mathrm{BM}0}=
    4.86\times10^{-2}\,,\quad
    \Omega_{\mathrm{R}0}=
    5\times10^{-4}
\end{equation}
are taken from the Planck mission data \cite{Planck2018parameters}. The initial (or, in fact,present day) conditions are taken in the most natural way $a_0=1, \dot{a}_0=H_0=1$, $\Phi_0$ is to be determined while $\dot{\Phi}_0=0$ in order to exclude stiff matter in the limit $\epsilon=0$. We also have to take into account the Hamiltonian constraints \eqref{m4} imposed on the data, cf. \eqref{InitCond}\footnote{Here, we work with the normalized scale factor $a_0=1$ and normalized cosmic time $T=H_{0}^{-1}=1$.}:
\begin{equation}\label{Cauchy2}
	1=\Omega_{\Lambda}+\Omega_{\mathrm{R}0}+\Omega_{\mathrm{BM}0}\Phi_{0}^{\gamma}\,,\end{equation}
that relates $\Omega_\Lambda, \Phi_0\ \mathrm{and}\ \gamma$.
Now, looking for the values, giving a good fit to the $\Lambda$-CDM model, we successfully find: 
  \begin{equation}\label{gamma value}
  	\Omega_{\Lambda}\approx 0.739\,,\gamma\approx 0.2450 \,,\Phi_0=946.507\,.
  \end{equation} 
The quality of the fitting is shown on Figs. \ref{fig:Scale-factor-near-1} and \ref{fig:Hubble-parameter-general-difference}, where $\Lambda$-CDM plot is based on Planck data \cite{Planck2018parameters}.

This good accuracy obeys Pantheon supernovae data and extends beyond the current epoch. In both cases, as expected, the parameters values are independent of $\epsilon$.
It should be also noted that $\Omega_\Lambda$ is not much different from Planck value $\Omega_{\Lambda\mathrm{Planck}}=0.6847$.
Accordingly, the calculated value of dust to baryonic matter ratio is also not much different from the Planck one:
  \begin{equation}\label{ratio}
  	\frac{\rho_{\mathrm{dust}}}{\rho_{\mathrm{BM0}}}=\Phi_0^\gamma=5.3598\,.
  \end{equation}
Moreover, the baryonic matter density:
\begin{equation}\label{BM conserved}
    \rho_{\mathrm{BM}}\equiv \rho_{\mathrm{BM0}}\,a^{-3}
\end{equation}
is conserved while the total dust density:
\begin{equation}\label{dust}
    \rho_{\mathrm{dust}}\equiv\rho_{\mathrm{BM0}}\,\Phi^{\gamma}\,a^{-3}
\end{equation}
 is not conserved and satisfies the equation:
\begin{equation}\label{dust not}
	\dot{\rho}_{\mathrm{dust}}+3H\rho_{\mathrm{dust}}=\gamma\frac{\dot{\Phi}}{\Phi}\,\rho_{\mathrm{dust}}.
\end{equation}
This makes the \textit{chameleon dark matter}:
\begin{equation}\label{dm chameleon}
    \rho_{\mathrm{DM}}\equiv\rho_{\mathrm{dust}}-\rho_{\mathrm{BM0}}
\end{equation}
an \textit{emergent quantity} whose density is \textit{not conserved}.

\begin{figure}[h]
	\centering
	\includegraphics[width=1\linewidth]{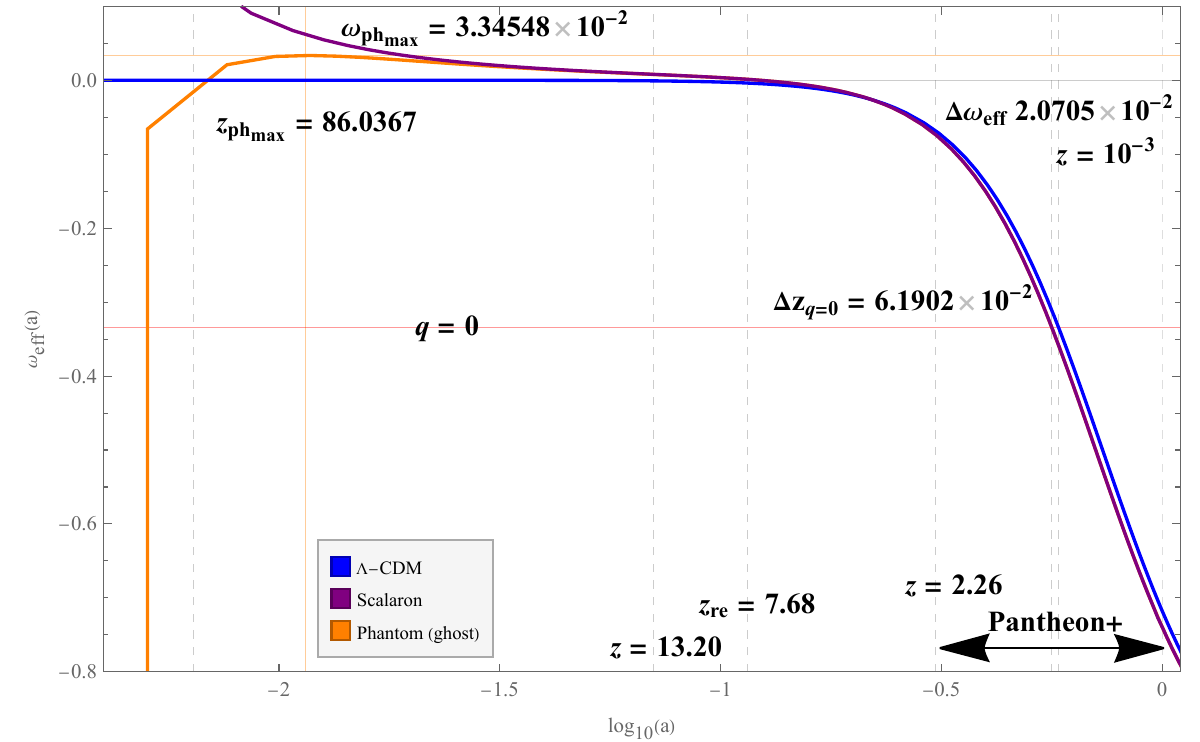}
	\caption{\scriptsize{Evolutions of the effective equation of state   \eqref{omega effective}.The differences among models seem to be insignificant from this perspective.}}
	\label{fig:omega-effective-comparison}
\end{figure}

Furthermore, using effective density and pressure \eqref{ab3} 
one can easily define an expression describing the  evolution of effective EoS  parameter, c.f. \eqref{ab1}, \eqref{eq1ab}:
\begin{equation}\label{omega effective}
	\omega_{\mathrm{eff}}(t)\equiv \frac{p_{\mathrm{eff}}(t)}{\rho_{\mathrm{eff}}(t)}=-\frac{2}{3}\frac{\dot{H}}{H^2}-1=\frac{1}{3}(2q-1)\,,
\end{equation}
where $q=-\frac{\dot{H}}{H^2}-1$ is a deceleration parameter. This is shown on  Fig.~\ref{fig:omega-effective-comparison}. $q=0$ corresponds to $\omega_{\mathrm{eff}}=-{1\over 3}$ that means transition from deceleration to acceleration era.\\
The evolution of this parameter also suggests a consistency with recent   Planck mission data, for which a \textit{mid-point redshift of reionization} has been estimated to be $z_{\mathrm{re}}=7.68\pm 0.79$ \cite{Planck2018parameters}. This compatibility leaves the door open to the possibility of creating a large-scale structure in such  models (keeping in mind the necessity to carry out a perturbation analysis).

\subsection{Small scale analysis: $t< 10^{-3}T_0$}

One of the most important aspects of cosmological models is their behavior during the earliest periods of time, if such initial time 
exists at all.
Solving the equations of motion \eqref{eq1m}-\eqref{eq2m}, we get, in fact, two scenarios which appear in much shorter scale still accessible for numerical calculations. Now, the "fine tuned"  value of  $\gamma=0.24500002$ parameter with a larger number of significant digits is needed in order to unveil these effects\footnote{For a smaller number of significant digits, the evolution period  to the value $a\sim 0$ (EFSTG) is slightly increased. This makes a comparison more tricky.}.
 
\begin{figure}[h]
	       \centering
	       \includegraphics[width=1\linewidth]{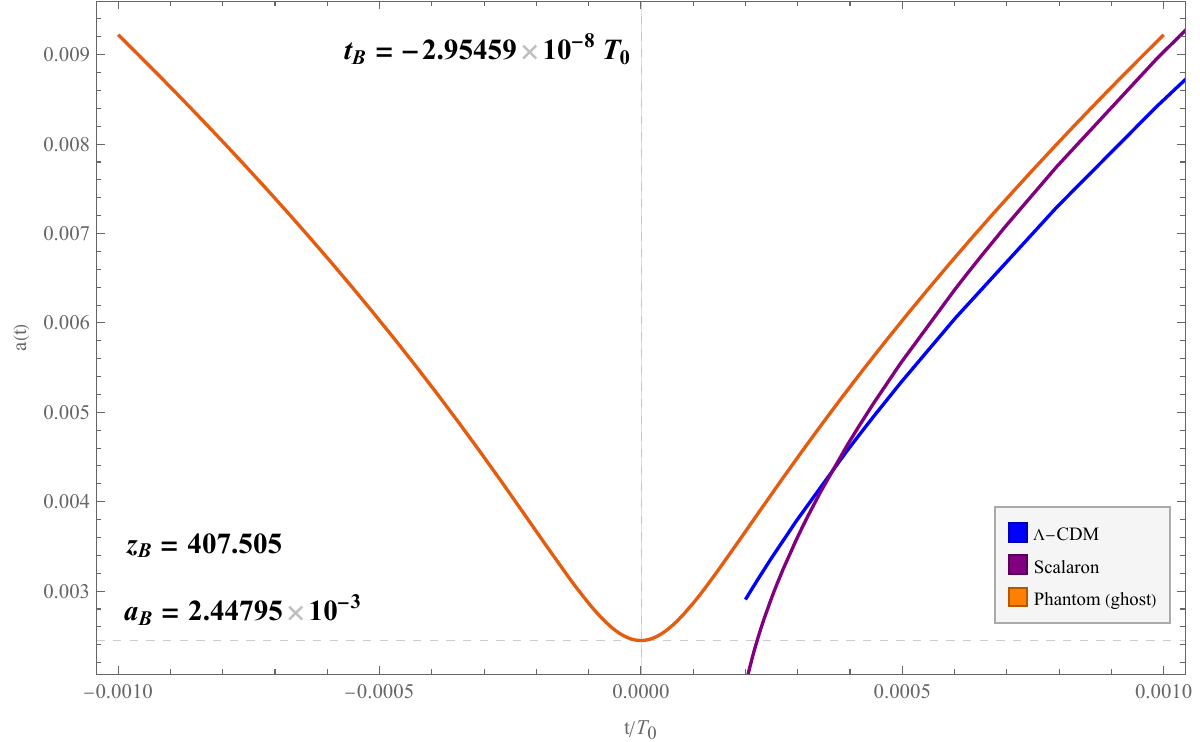}
	       \caption{\scriptsize{Behavior of the scale factors nearby a LCDM BB initial singularity: standard (with a singularity) for EFSTS and non-standard (with a bounce) for EFSTG.}}
	       \label{fig:Scale-factor-near-bounce}
    \end{figure}

In the first scenario, for $\epsilon=1$, we get a model with an initial singularity (in fact, very similar to the scenario considered in \cite{Pogosian2023}). In the case of $\epsilon=-1$ we are dealing with the existence of a phase of (plausible non-singular) \textit{cosmological bounce} (Fig.~\ref{fig:Scale-factor-near-bounce}). This phenomenon occurs in models related to the so-called Bounce Cosmology\footnote{For review papers on this subject, see \cite{Novello2008,Brandenberger2017,Battefeld2015}.}.

\begin{figure}[h]
	\centering
	\includegraphics[width=1\linewidth]{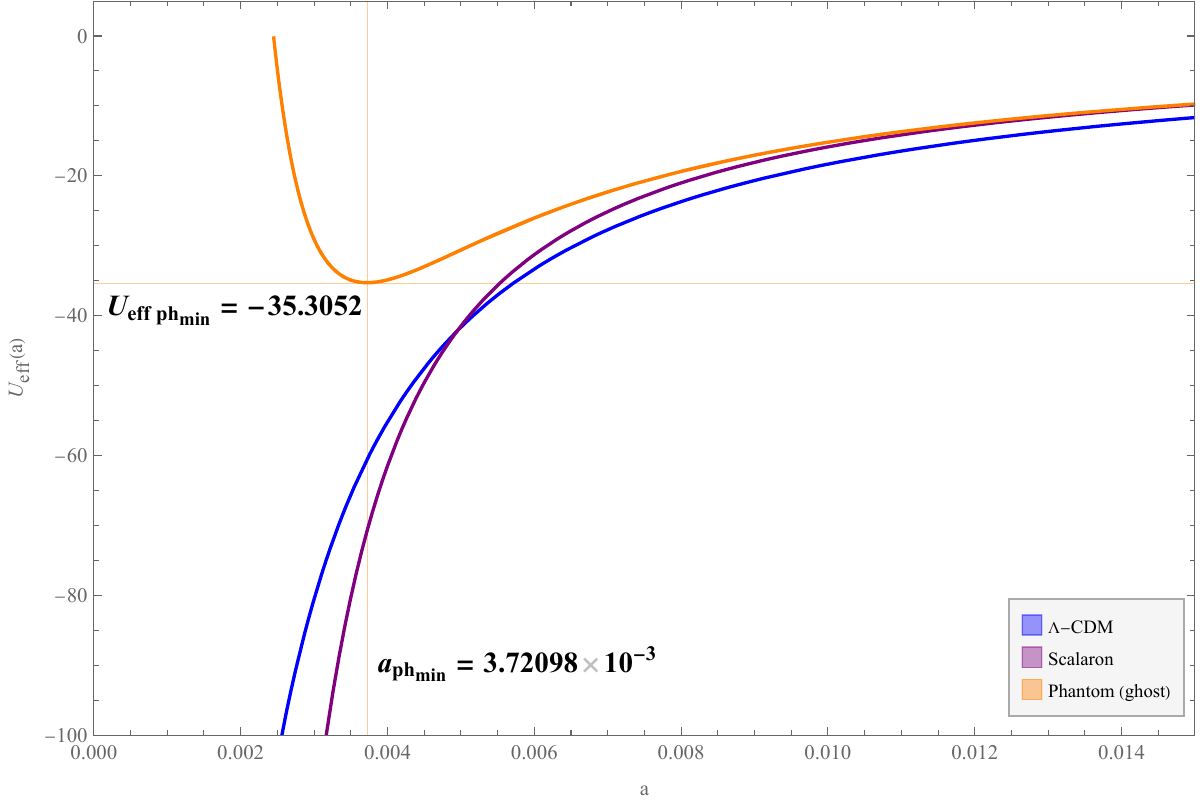}
	\caption{\scriptsize{Evolution of the effective potential as a function of the size of the Universe (scale factor) in the cosmological time interval: $-0.02 T_0 \leq t\leq 0.02 T_0$ ($\epsilon=-1$), $2\times 10^{-4} T_0 \leq t\leq 0.02 T_0$ ($\epsilon=1$) and $0 < t\leq 0.02 T_0$ ($\Lambda$-CDM).}}
	\label{fig:Ueff-near-a-equal-0-comparison}
\end{figure}

In the case of the EFSTS model, we have a classic example of a model of the $\Lambda$-CDM type, i.e., the case where a necessary aspect of solving the initial singularity problem is to implement an external mechanism (e.g. cosmological inflation, quantum gravity corrections). On the other hand, the EFSTG model is an example of the so-called \textit{matter bounce} scenario \cite{Brandenberger2020}, in which it is the exotic matter (in such a case, the kinetic energy of the phantom field) that leads to the contraction phase. Then a bounce occurs, followed by the period of accelerated expansion of the Universe with the presence of possible reheating epoch (see, Fig.~\ref{fig:Ueff-near-a-equal-0-comparison} and description below it).

    In the later epochs of the Universe evolution, the models show pretty good agreement with the reference model, i.e. the magnitude of $H(t)$ is a decreasing function aiming at a constant positive value (Fig.~\ref{fig:Hubble-parameter-general-difference}), which can be regarded as the era of the cosmological constant dominance, for which the value of the Hubble parameter is constant over time.

\begin{figure}[h]
	       \centering
	       \includegraphics[width=1\linewidth]{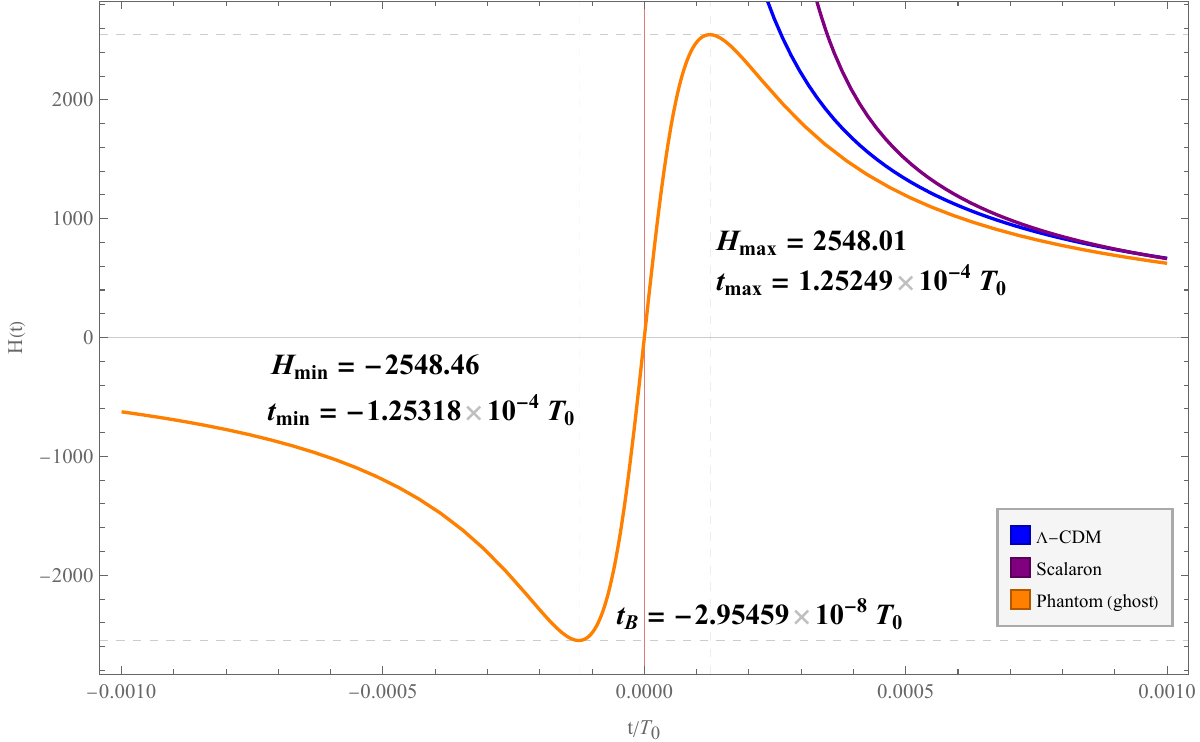}
	       \caption{\scriptsize{Behavior of the Hubble parameters nearby a LCDM BB initial singularity. One sees non-singular evolution for $\epsilon=-1$ case.}}
	       \label{fig:Hubble-parameter-near-bounce}
    \end{figure}

    In contrast, when it comes to the behaviour of the Hubble parameter near $t=0$, the EFSTG model is more challenging. That case involves a \textit{matter bounce} scenario \cite{Brandenberger2020}. The phase of cosmological bounce is illustrated in Fig.~\ref{fig:Hubble-parameter-near-bounce}.

    The Hubble parameter slowly decreases in value until it reaches a minimum, then passes through a value equal to zero (bounce moment). The last stage is to reach a maximum value and gradually decrease in value. This is standard behavior for models with matter-dominated cosmological bounce \cite{Brandenberger2020}.

\begin{figure}[h]
	       \centering
	       \includegraphics[width=1\linewidth]{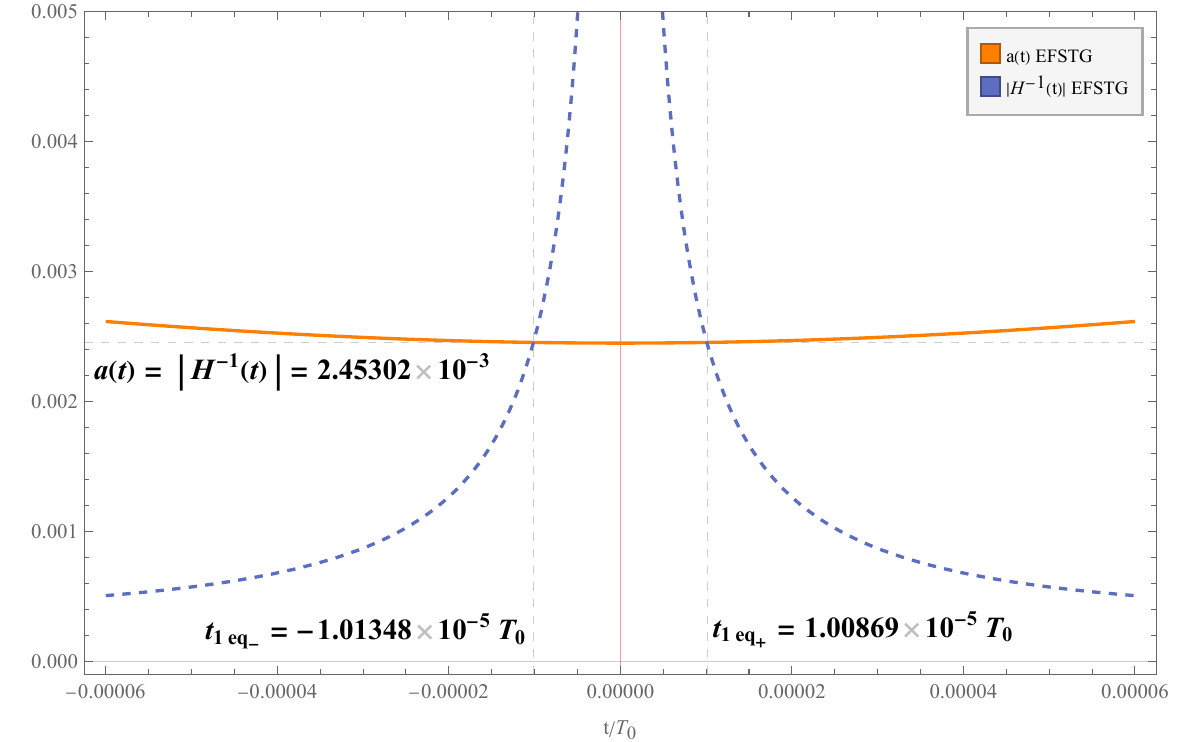}
	       \caption{\scriptsize{Hubble horizon for the EFSTG model. The scale factor entering below the horizon at moments marked on the graph represents the wavelength of the fluctuation mode - this is typical behavior for the matter bounce phase \cite{Brandenberger2012}.}}
	       \label{fig:Hubble-horizon-near-bounce}
    \end{figure}

    At this point, we can also consider an important quantity that can indicate the nature of the bounce, namely the Hubble horizon, which is the inverse of the Hubble function: $R_{\mathrm{H}}=H^{-1}(t)$. In Fig.~\ref{fig:Hubble-horizon-near-bounce} one can see that wavelength of the fluctuation mode represents by a scale factor function (i.e., $\lambda\propto a$) enters under the horizon, shortly before as well as exits shortly after the bounce. According to our numerical solutions, this phase lasts about 279,000 years.

\begin{figure}[h]
		\centering
		\includegraphics[width=1\linewidth]{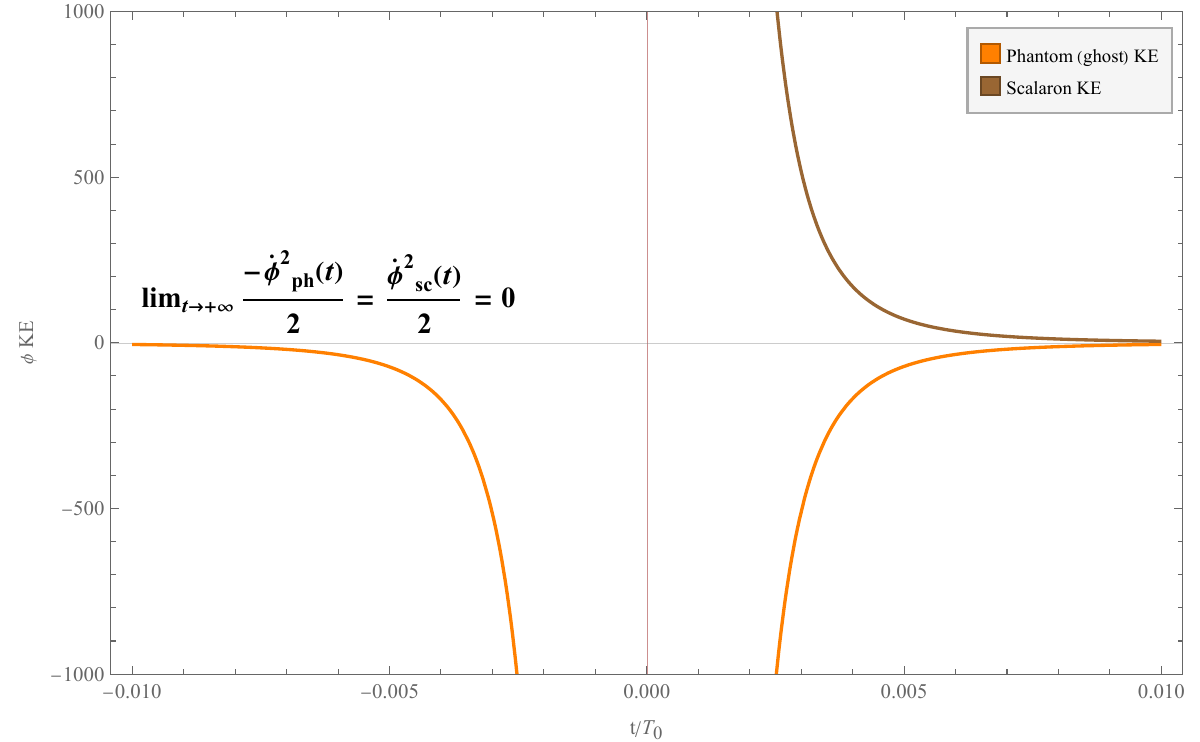}
		\caption{\scriptsize{Kinetic energy of the scalar field as a decisive factor for the earliest stages of the evolution of the Universe. In the case of the scalaron field, it shows a positive character as for the canonical scalar field and a negative character typical for the ghost fields.}}
		\label{fig:phi-KE-near-bounce}
	\end{figure}

    As was mentioned earlier, the transition phase between contraction and expansion (or by contraction in the case of EFSTS) for the EFSTG model is dominated by the kinetic energy of the ghost field. As shown in Fig.~\ref{fig:phi-KE-near-bounce}, the kinetic energy of the scalaron field is a decreasing function in time asymptotically going to zero. In contrast, in the second model we are considering, we are dealing (most likely) with the singular nature of the kinetic energy of the field at the very moment of cosmic bounce. This energy also asymptotically tends to zero, however, in such a case in the limit with $t\to\pm T_0$.

    The most significant differences are seen in the early periods of the Universe evolution. In contrast to the LCDM model, positive maximum were obtained for the model with $\epsilon=-1$, while for the scalaron field model the parameter of the equation of state seems to have no maximum but in the early period also exhibits a positive-value character (again, Fig.~\ref{fig:omega-effective-comparison}). For both models, the behavior of the equations of state converges to LCDM-like early in the evolution of the Universe (about 13.8 mln years after the Big Bang/Bounce).

\begin{figure}[h]
	       \centering
	       \includegraphics[width=1\linewidth]{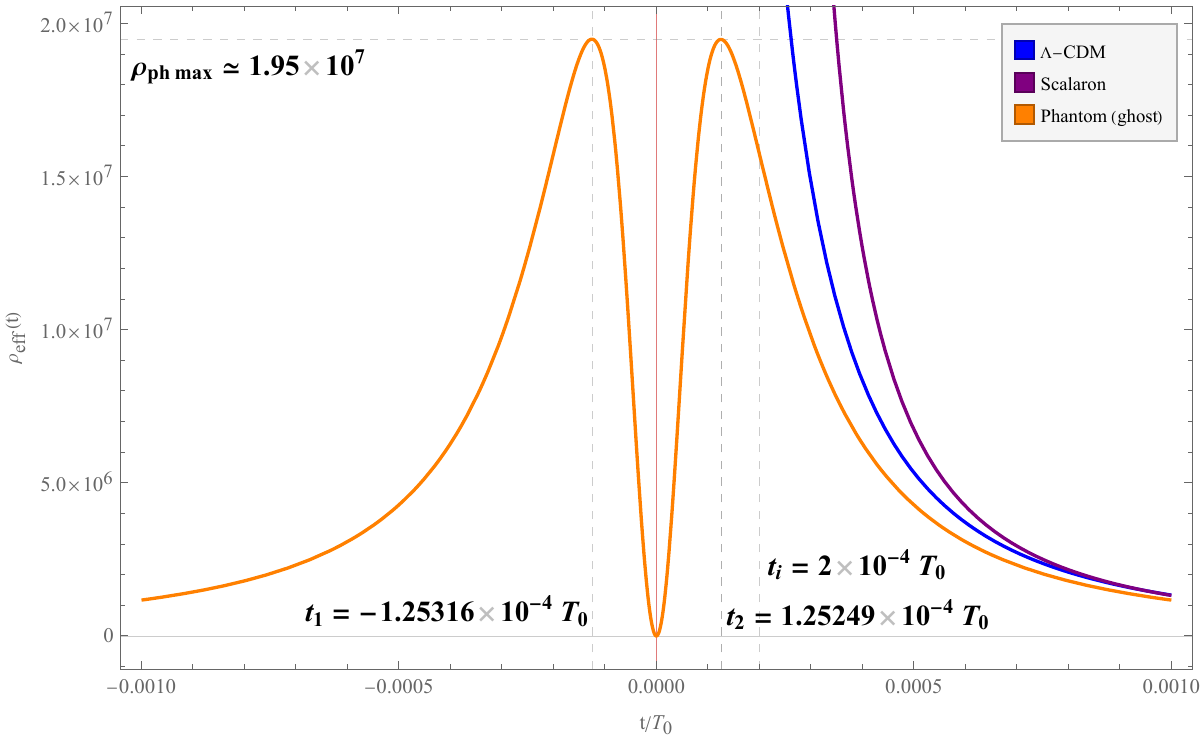}
	       \caption{\scriptsize{The typical evolution of the energy density function \eqref{omega effective} for models with an initial singularity (EFSTS) and the evolution with two maxima for a model with a bounce phase (EFSTG) \cite{Brandenberger2020}.}}
	       \label{fig:rho-comparison-near-bounce}
    \end{figure}

    Also in the case of effective energy density concept \eqref{omega effective}, one can see significant differences in the earliest period of the evolution of the Universe. The scalaron field model deviates slightly from the case of the reference model, while the ghost field model reveals the tangible presence of the epoch of dominance of non-standard matter (kinetic energy of the phantom field) (Fig.~\ref{fig:rho-comparison-near-bounce}).

    The maximum of the energy density before the bounce corresponds to the moment when the effective negative density of the ghost (stiff matter) field turns on, during the slow contraction the effective density decreases (probably) to zero (the moment of the bounce) and then increases to a new maximum of the same value as before. When the kinetic energy of the phantom field becomes no longer dominant we return to standard evolution, i.e. the energy density decreases as the size of the Universe increases.
    
As in the case of the reference model ($\Lambda$-CDM), the approximated influence of individual components on the global evolution can be represented in a simple way by formulating the effective (Newtonian) potential. Such potential, is expressed, in the case of our models, as follows:
\begin{equation}\label{U eff}
	U_{\mathrm{eff}}\left(a\right)=-\frac{1}{2}\left(\frac{\epsilon}{6}\dot{\Phi}^2 a^2+\Omega_{\Lambda}a^2+\Omega_{\mathrm{R}}a^{-2}+\Omega_{\mathrm{BM}}a^{-1}\Phi^{\gamma}\right).
\end{equation}
With the help of Fig.~\ref{fig:Ueff-near-a-equal-0-comparison}, a significant conclusion can be drawn. Namely, the local minimum of the Newtonian-type potential for the EFSTG model could represent the \textit{reheating} phase of the Universe evolution. This takes place in our case for a redshift approximately equal to $z=267.75$ what corresponds to the epoch of large-scale structure formation. However, in the case of the model with $\epsilon=-1$, it should be borne in mind that the scenarios proposed so far for obtaining the large-scale structure of the Universe are based on the condensation effect (the Universe obtains a critical temperature), in which structures are formed "immediately" and not as a result of a long-term process at time \cite{Matos2000SFDM,Matos2000,Matos2001}\footnote{Examples of such particles include axions and axion-like particles (ALPs) \cite{Marsh2016,Kuster2008}.}.

The non-standard new segment in the expression \eqref{U eff} is a component that depends on the type of kinetic energy of the scalar field (canonical or ghost) mimicking dark matter. The crucial characteristic of stiff matter\footnote{More details on stiff matter cosmology can be found at \cite{Chavanis2015stiff}.} is that its energy density dilutes more slowly as the universe expands and thickens more slowly  as the universe contracts compared to other forms of matter/energy.

An unusual aspect of the effective potential for the EFSTG model is the presence of a positive contribution from the kinetic energy of the scalar field. This results in the possibility of contraction and then, as a result of the decay of part of the energy of the scalar field, accelerated expansion of the Universe. This process can be regarded as the presence of a reheating mechanism in models that are effective approximations of the unknown (now or never?) epoch of quantum gravity. It may also indicate the possibility of the existence of currently unknown states of matter/types of energy in the Universe.

\subsection{Relevant cosmological aspects regarding scalaron and phantom (ghost) scalar field}\label{Phantom intro}

Phantom energy characterized by the equation of state parameter $\omega<-1$ appears in individual braneworld models \cite{Sahni2003,Sahni2012} or Brans-Dicke theory \cite{Huang2007}. The simplest possible approach to get this effect is to introduce a ghost\footnote{For a more general discussion from QFT perspective, see \cite{Hawking2002}.} scalar field with a negative kinetic term in the action \cite{Caldwell2002}. Such a scalar field naturally appears in effective theories originated from type IIA String Theory \cite{IIA1,IIA2} and  low-energy limit of F-theory reformulated in 12-D type IIB action \cite{Ftheory}.

\begin{figure}[h]
	       \centering
	       \includegraphics[width=1\linewidth]{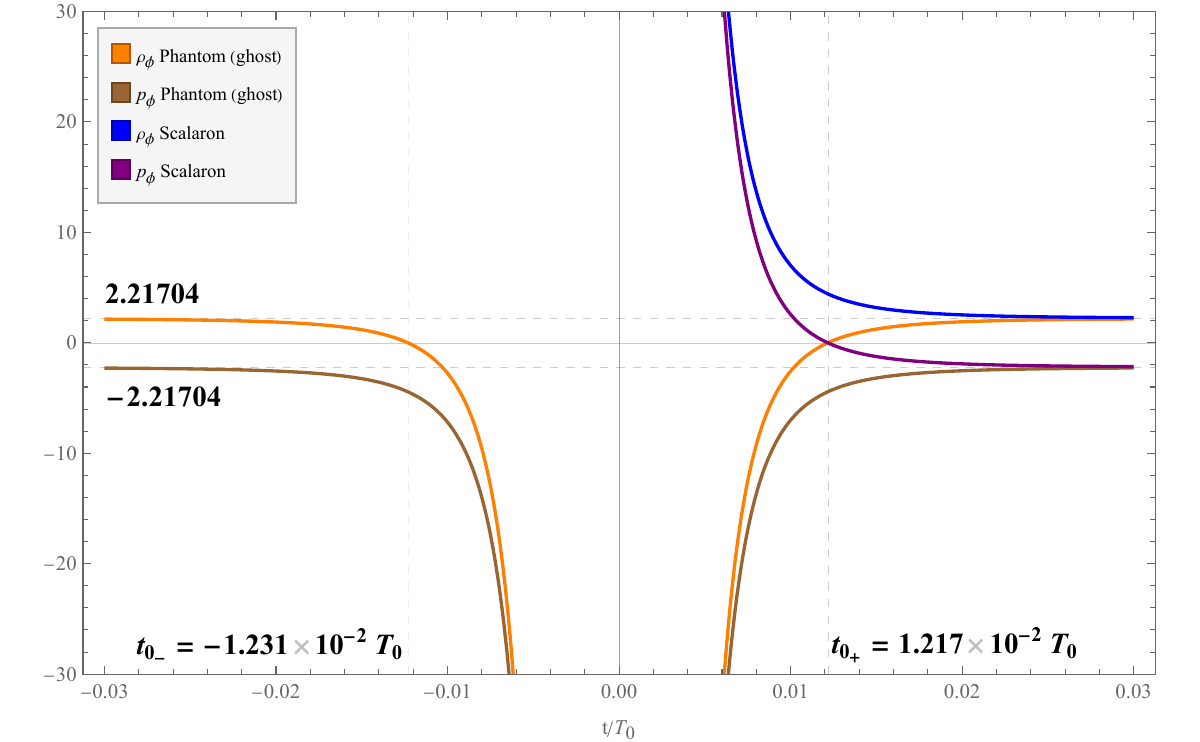}
	       \caption{\scriptsize{In the case of the matter bounce phase, the value of the energy density and pressure exhibited by the phantom field (\eqref{omega phi parameter}) tends to a constant (finite) negative value. In contrast, for the scalaron field (also \eqref{omega phi parameter}), we observe a decrease in both values over time. The values of energy density and pressure for both scalar fields tend asymptotically to the same values.}}
	       \label{fig:p-phi-rho-phi-sc-ph}
    \end{figure}

The phantom (ghost) field in cosmology shows quite interesting properties, e.g. its value of the energy density $\rho_{\Phi}$ increases with time in the period after cosmological bounce (Fig.~\ref{fig:p-phi-rho-phi-sc-ph}), the speed of sound equals the speed of light (although there are models with subluminal values of this speed, for instance \cite{Armendariz1999,Parker2001,Superq}), there is also a correlation between such types of fields and the de Sitter-CFT \cite{McInnes2002}.

Alternative cosmological models often explore the introduction of phantom or ghost fields to explain observations that deviate from the predictions of the standard (LCDM) cosmological model. One such intriguing proposal involves ultra-light or massless axions/ALPs, which can significantly influence cosmological dynamics. A notable example is the \textit{photon-axion conversion mechanism}\footnote{For more details on this topic, see \cite{Mirizzi2008}}, which, when combined with a DE model characterized by an EoS parameter $\omega\gtrsim-1$, can mimic a cosmic EoS as extreme as $\omega\simeq-3/2$ \cite{Csaki2005}. While current observational data do not necessitate such an extreme EoS\footnote{It should be noted that in the case of a time-dependent EoS, $\omega(a)=\omega_{0}+(1-a)\omega_{a}$, the DESI data tend to favour solutions with $\omega_{0}>-1$ and $\omega_{a}<0$. Furthermore, when combining DESI and CMB data, this yields $\omega_{0}=-0.45^{+0.34}_{-0.21}$ and $\omega_{a}=-1.79^{+0.48}_{-1.00}$, indicating a discrepancy of approximately $2.2\sigma$ from the $\Lambda$-CDM model \cite{DESI2024}.}, the possibility remains an important consideration. This is particularly relevant given that phantom and ghost field models, despite their ability to produce such exotic states, often challenge foundational principles of QFT and GR. Thus, exploring axion/ALP-based mechanisms offers a promising alternative that avoids these theoretical pitfalls while still accommodating potential deviations in cosmic expansion dynamics.

    The fundamental difference in the description of the two fields we are considering is the different form of the equation of state describing the evolution of the scalaron and phantom (ghost). The scalaron field involves the standard form known from the canonical scalar field scenario in cosmology (i.e. $\epsilon=1$). This, in the case of our generalization, can be described as follows:
    \begin{equation}\label{omega phi parameter}
        \omega_{\Phi}^{(\epsilon)}\equiv\frac{p_{\Phi}}{\rho_{\Phi}}=\frac{\frac{1}{2}\left(\epsilon\dot{\Phi}^2-V(\Phi)\right)}{\frac{1}{2}\left(\epsilon\dot{\Phi}^2+V(\Phi)\right)}=1-\frac{2V(\Phi)}{\epsilon\dot{\Phi}^2+V(\Phi)}.
    \end{equation}
    On the other hand, for the "non-canonical" ghost field, we are dealing with a specific form of the equation \eqref{omega phi parameter} introducing the changed (negative sign) up front of the kinetic term, namely $\epsilon=-1$. One can see from the expression \eqref{omega phi parameter} that the equation of state for the phantom (ghost) field is indeterminate when the kinetic energy of the scalar field is equal to its potential energy (cosmological constant).

\begin{figure}[h]
	\centering
	\includegraphics[width=1\linewidth]{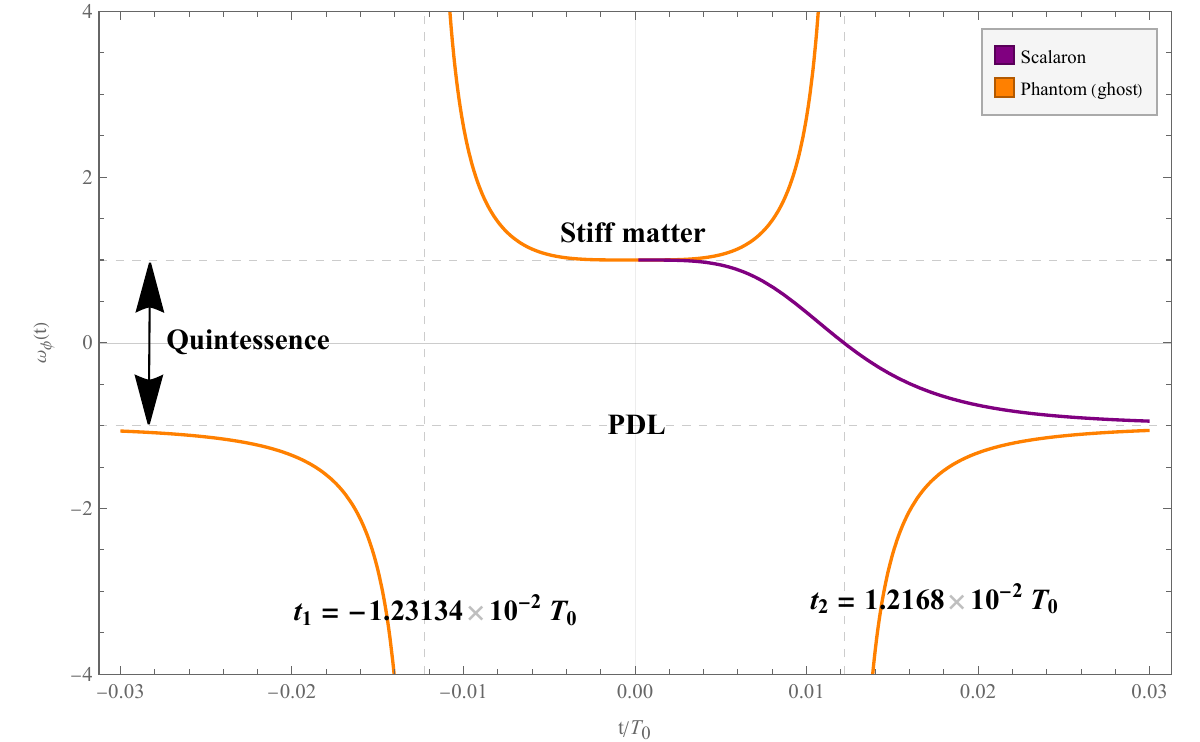}
	\caption{\scriptsize{The scalar field equation of state parameter \eqref{omega phi parameter} as a criterion to characterize the different effective (matter) phases of the $\Phi$ (DM) behavior. PDL stands for Phantom Divide Line.}}
	\label{fig:omega-phi-near-bounce}
    \end{figure}

    Most likely, at this point, a kind of phase transition occurs and the dark matter described by this field goes from effective stiff matter in the epoch of the matter bounce to an effective phantom phase asymptotically transformed into an effective factor incorporated into dark energy (Fig.~\ref{fig:omega-phi-near-bounce}). Such behaviour exhibited by the scalar field may indicate a potential shift in the sign of the 'effective' cosmological constant/dark energy \cite{Akarsu2020}, which could be a promising avenue for resolving the Hubble tension \cite{Akarsu2021}.

    In the case of the scalaron field, the situation is different. In the earliest period of the evolution, the field also exhibits behavior that characterizes stiff matter ($\omega_{\Phi}^{\mathrm{(sc)}}\simeq1$), but then goes through a dust epoch ($\omega_{\Phi}^{\mathrm{(sc)}}\simeq0$) and then mimics the behavior as for a cosmological constant ($\omega_{\Phi}^{\mathrm{(sc)}}\simeq-1$). Thus, it may also represent a kind of \textit{quintessence} field known from the attempt to unify dark matter and dark energy \cite{Tsujikawa2013} (again Fig.~\ref{fig:omega-phi-near-bounce}).

    By considering the evolution of the equation of state for the obtained models with a scalar field depending on the scale factor, it is possible to discuss the problem of observations of galaxies with a high-redshift value, which pose a challenge for the currently recognized theory of the large-scale structure formation of the Universe (alternatively, it may be a contribution to the discussion regarding the age of the Universe/duration of the epoch after a possible cosmological bounce such as in EFSTG model).
    \begin{figure}[h]
	\centering
	\includegraphics[width=1\linewidth]{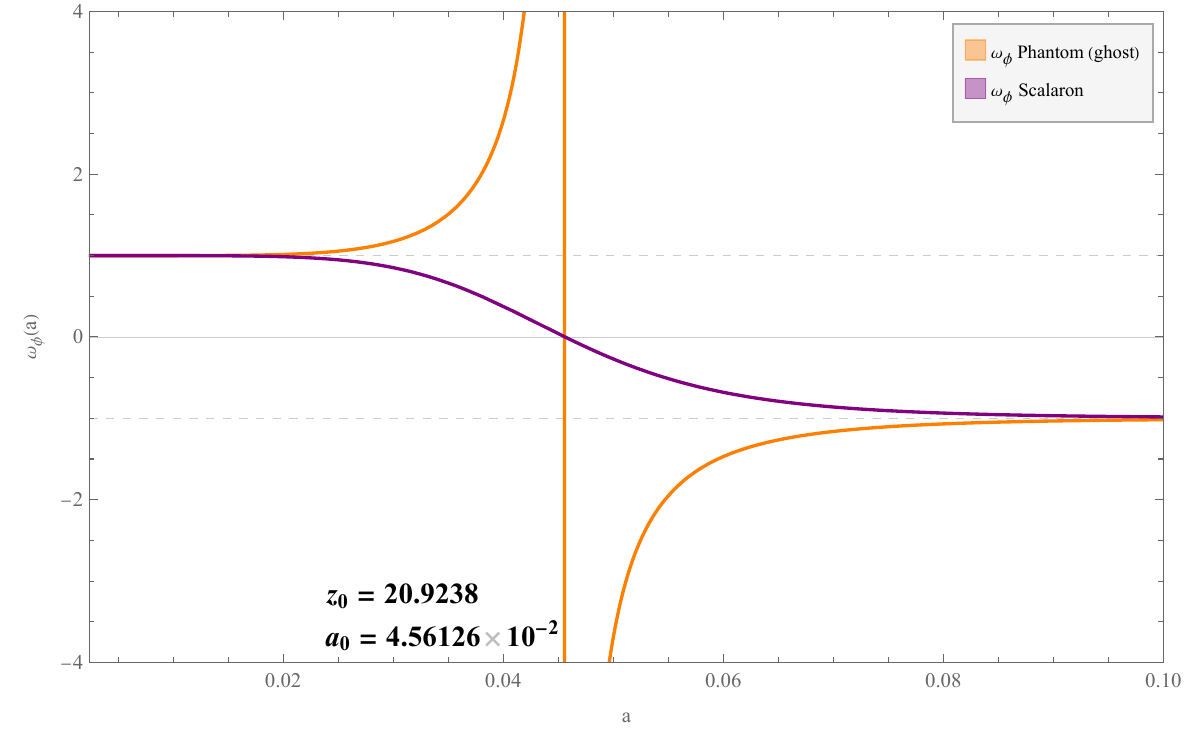}
	\caption{\scriptsize{Dependence of the $\omega_{\Phi}$ parameter versus the scale factor with an apparent transition between positive and negative values of the equation of state \eqref{omega phi parameter}.}}
	\label{fig:omega-phi-zero-a}
    \end{figure}

    As one can see in Fig.~\ref{fig:omega-phi-zero-a}, the transition between the negative and positive value of the $\omega_{\Phi}$ parameter (for both models) occurs for the redshift of $z_{0}=20.9238$ (about 168 mln years after the Big Bang/Bounce). Possible future observational data from the JWST may confirm or dismiss this concept.

    However, one should also remember the potential problems associated with the phantom (ghost) scalar field. One of the main aspects is the so-called \textit{UV instability}. To this group, we can include vacuum instability problems due to particle production caused by the lack of a minimum energy density $\rho_{\Phi}$ constraint, and the presence of a \textit{MeV cut-off} to avoid producing excess gamma radiation \cite{Copeland2006}.

    \subsection{Jordan frame evolution of the models}
    Our model was proposed originally in the Einstein frame \eqref{Einstein frame action} as a special case of the most general ST action \eqref{general action} parametrized by:
    \begin{equation}\label{EF parametrization}
    	\begin{split}
    	& \ca(\Phi)=1, \qquad \cb(\Phi)=\epsilon\equiv\pm 1, \\
        & \cv(\Phi)=\cv_{\mathrm{DE}}\equiv 2\Lambda, \quad\alpha(\Phi)=\gamma\ln{\Phi}\,.
    	\end{split}
    \end{equation}
   It can be  also analyzed in the Jordan frame with the reparametrization of the following form (see, \cite{Flanagan2004,Jarv,Borowiec2021}):
    \begin{equation}\label{JF parametrization}
        \begin{dcases}
            \bar\ca(\Phi)=\Phi^{-2\gamma}\equiv\Psi, \\
            \bar\cb(\Phi)=\Phi^{-2\gamma}\left(\epsilon-6\gamma^2\Phi^{-2}\right)=\Psi^{-\frac{1+\gamma}{\gamma}}\left(\frac{\epsilon}{4\gamma^2}-{3\over 2}\Psi^{1/\gamma}\right), \\
            \bar\alpha(\Phi)=0, \\
            \bar\cv(\Phi)=\Phi^{-4\gamma} V_{DE}=\Psi^{2} V_{DE}.
        \end{dcases}
    \end{equation}
Instead of a non-minimal coupling to the matter, we have a non-minimal coupling to gravity. The matter stress-energy tensor is conserved now and the conformally transformed Jordan frame metric becomes:
\begin{equation}
\bar g_{\mu\nu}= \Phi^{2\gamma} g_{\mu\nu}	=\Psi^{-2}  g_{\mu\nu}\,,
\end{equation}
which does not distinguish between baryonic and dark matter, is a solution of the corresponding field equations \eqref{STT general eq} with modified frame functions.\\
In this sense, one says that they are mathematically equivalent. We should also notice that redefinition of the scalar field: $\Phi\mapsto \Psi$ does not change the scale factor evolution.  
    Cosmological time and the scale factor in the Jordan frame are expressed through the corresponding quantities in the Einstein frame:
    \begin{equation}\label{units}
    	\begin{dcases}
    		\bar{a}=e^{\alpha(\Phi)} a, \\
    		d\bar{t}=e^{\alpha(\Phi)} dt.
    	\end{dcases}
    \end{equation}
    Consequently, the Jordan frame Hubble parameter is of the form:
    \begin{equation}\label{H Jordan}
    	\bar{H}\equiv\frac{\frac{d\bar{a}}{d\bar{t}}}{\bar{a}}=e^{-\alpha(\Phi)}\left[H+\alpha'(\Phi)\dot{\Phi}\right].
    \end{equation}
    Therefore, the deceleration parameter in the Jordan frame takes the following form:
    \begin{equation}\label{q Jordan}
    \begin{split}
        \bar{q} 
        & = \frac{q H^2- \alpha' (H\dot{\Phi}+\ddot{\Phi} )-\alpha''\dot{\Phi}^2 }{\left(H+\alpha' \dot{\Phi}\right)^2}\,.
    \end{split}
    \end{equation}
Plugging now our numerical solution to the formula \eqref{H Jordan} one can check that Jordan frame evolution represents a different physical scenario than the previous one.
\subsubsection{Scalaron field model}
\begin{figure}[h]
	\centering
	\includegraphics[width=1\linewidth]{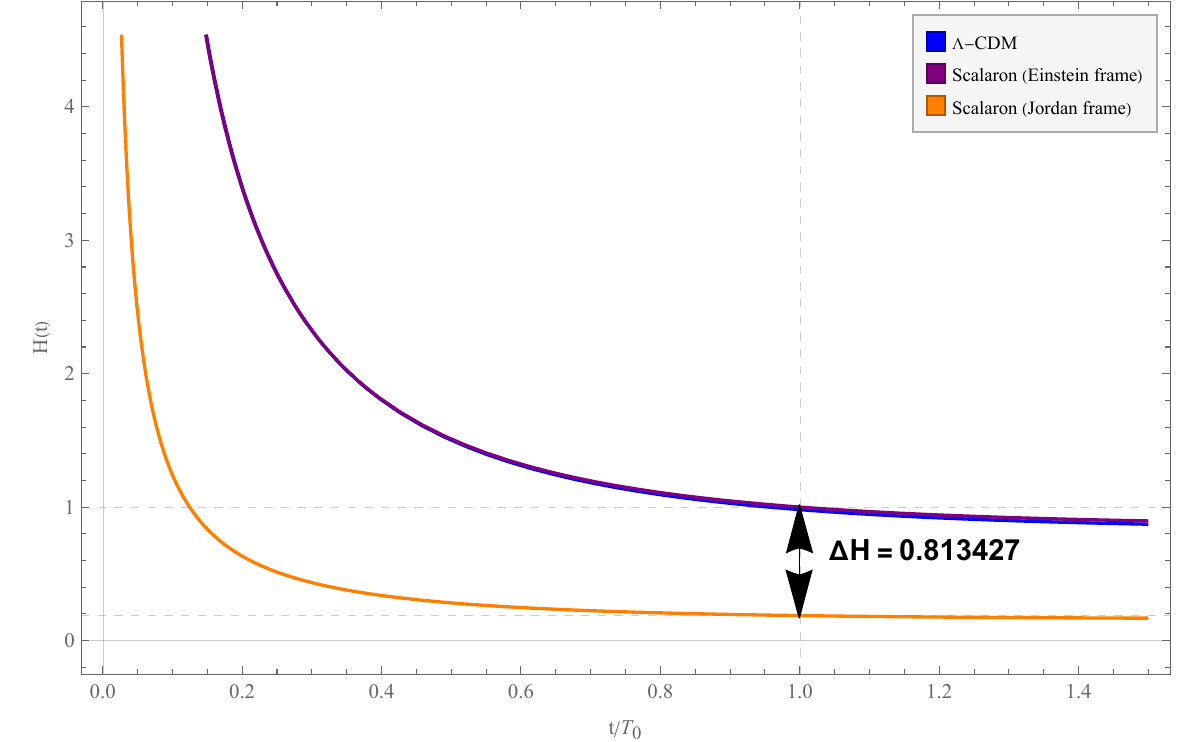}
	\caption{\scriptsize{Comparison of the evolution of the Hubble parameter in the case of scalaron field in the Einstein and Jordan \eqref{H Jordan} conformal frames.}}
	\label{fig:Comparison-H-Scalaron}
\end{figure}

Formula defined by the relation \eqref{H Jordan} allows us to determine the course of cosmic expansion in the model specified in the frame, in which we take into account the non-minimal coupling between the quintessence field and the matter contained in the Universe. Evolution of the Hubble parameter within such a frame is visualized in Fig.~\ref{fig:Comparison-H-Scalaron}.

The first evident observation arising from the course of the expansion of the Universe in a model containing a canonical scalar field is the obvious failure to satisfy LCDM-type boundary conditions (assumed in advance in the Einstein frame). In addition, there is a significant shift in the value of the Hubble parameter towards smaller values compared to the reference model.

Conclusively, we come to the self-evident conclusion that the formulation of our theory in two conformal frames \textit{does not lead} to the two physically equivalent models.

\subsubsection{Phantom (ghost) field model}
In the case of the phantom (ghost) field, we have a situation that is more interesting from a theoretical point of view. Namely, in this formulation of the theory, we are faced with a situation where our matter component represented by the kinetic energy of the scalar field takes on negative values of the effective energy density during the matter bounce phase.

However, in theoretical physics, possible realizations of such exotic energy densities are known regarding QFT, which allows to violate the WEC in the form of locally negative energy densities and fluxes \cite{Ford1992,Ford1993}, averaged energy conditions \cite{Ford1995}, and higher-dimensional theories such as Kaluza-Klein \cite{Witten1982}\footnote{Regarding this aspect, it is worth recalling at this point the so-called QEIs (Quantum energy inequalities), which state that an average taken along a timelike curve will produce a finite (but possible negative) lower bound on energy state (restricted in the magnitude and duration). For more details, see \cite{Kontou2020}.}.
\begin{figure}[h]
	\centering
	\includegraphics[width=1\linewidth]{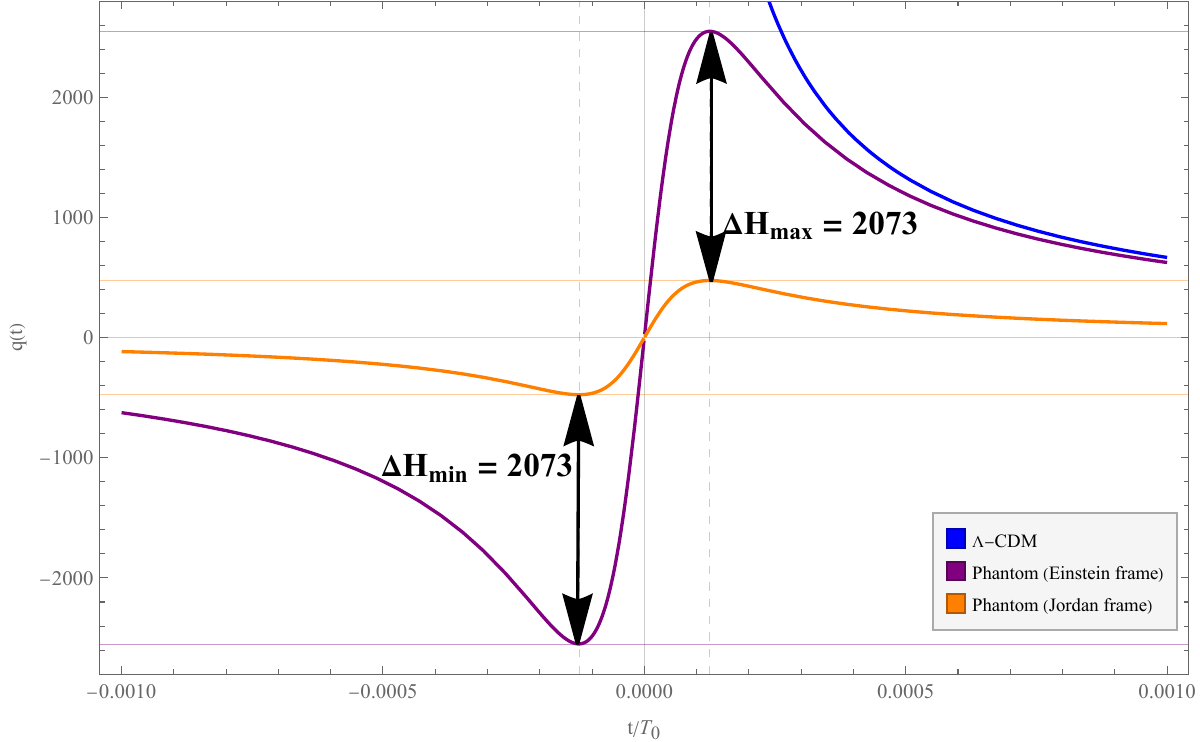}
	\caption{\scriptsize{The general structure of the evolution of $H(t)$ is preserved for both frames, however, the Jordan frame \eqref{H Jordan} shows a flattening of this curve.}}
	\label{fig:Comparison-H-Phantom}
\end{figure}
\begin{figure}[h]
	\centering
	\includegraphics[width=1\linewidth]{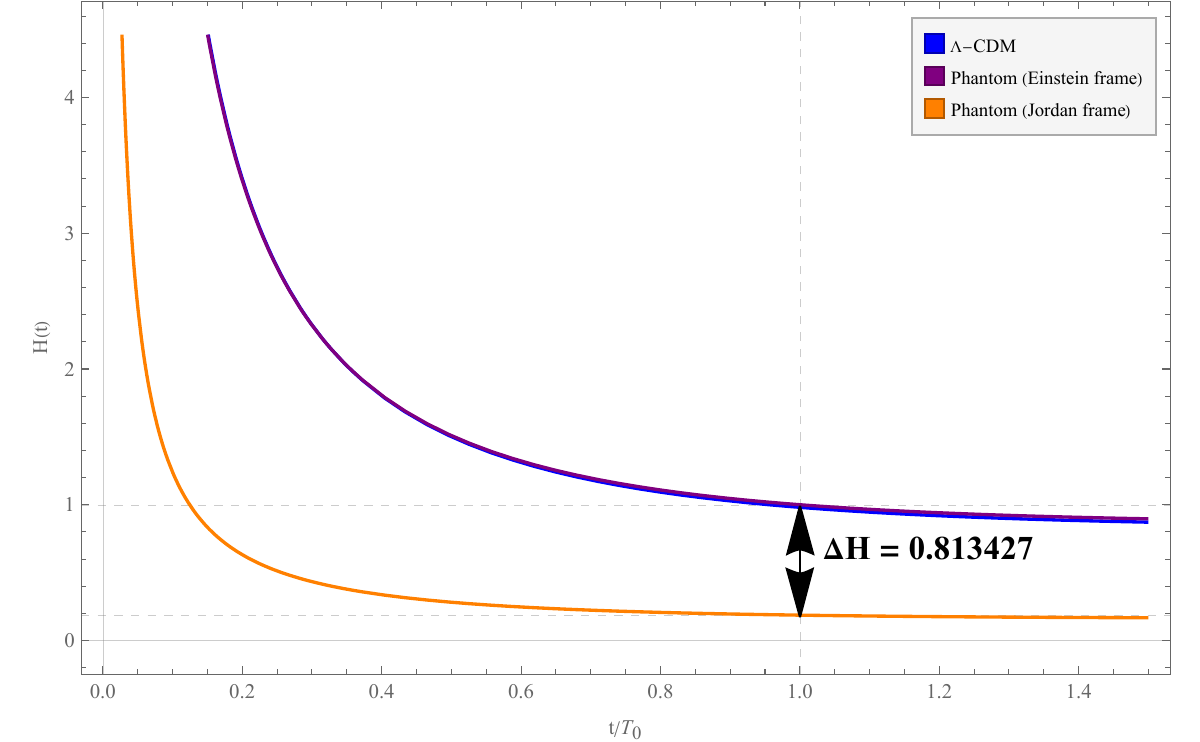}
	\caption{\scriptsize{The behavior of $H(t)$ \eqref{H Jordan} for the phantom model is analogous to that of the quintessence one.}}
	\label{fig:Comparison-H-Phantom-near-1}
\end{figure}

Fig.~\ref{fig:Comparison-H-Phantom} shows the evolution of $H(t)$ for the phantom model in the Jordan frame. Analogous to the case of the quintessence field, also here we have a breakdown of LCDM-like Cauchy data in new conformal frame. The amplitude with which the Hubble parameter evolves is significantly reduced. The value of the Hubble constant $H_0$ is also much smaller than in the corcondance model (Fig.~\ref{fig:Comparison-H-Phantom-near-1}).

It is worth noting, however, that the main structure characterizing the behavior of the Hubble parameter for the matter bounce phase is preserved (including the duration of this epoch).

Also with this model, \textit{we are not entitled to claim physical equivalence} between the formulation of the theory in different conformal frames.  It is also worth mentioning that similar models and conclusions to those considered in this paper, however, from the point of view of the Jordan frame were investigated using dynamical system techniques in \cite{Hrycyna2020,Hrycyna2021} and analytical ones in \cite{Capozziello2010}.

\subsection{PPN constraints}
The \textit{Parametrized Post-Newtonian} (PPN) formalism represents a theoretical framework utilized to describe and compare the predictions of disparate gravitational theories within the context of the weak-field, slow-motion regime. This regime is characterized by the presence of weak gravitational fields and velocities of bodies that are significantly less than the speed of light. The formalism expands the metric tensor in a series of terms that represent corrections to Newtonian gravity, each parameterised by a set of dimensionless coefficients known as PPN parameters \cite{Will2014,Will2018}. These PPN parameters quantify various potential deviations from General Relativity (GR), thus allowing for a systematic comparison of alternative gravitational theories such as ST models \cite{Coc2006}.

The analysis presented here investigates the compatibility of PPN parameters, employing a notation that is drawn from \cite{Coc2006}. In the context of our particular case, this implies that\footnote{According to our convention: $\Phi=\sqrt{2}\varphi_{*}$.}:
    \begin{equation}
        A\left(\varphi_{*}\right)=e^{\alpha\left(\varphi_{*}\right)}=
        2^\frac{\gamma}{2}\varphi_{*}^{\gamma}\equiv \Phi^{\gamma}.
    \end{equation}
In such a case the strength of the coupling between the SF and the matter fields is quantified by the following parameters \cite{Coc2006}:
    \begin{equation}
        \alpha_{\mathrm{PPN}}\left(\varphi_{*}\right)\equiv\frac{d}{d\varphi_{*}}\left[\ln{\left(e^{\alpha\left(\varphi_{*}\right)}\right)}\right]=\alpha'\left(\varphi_{*}\right)=\frac{\gamma}{\varphi_{*}}=\frac{\sqrt{2}\gamma}{\Phi},
    \end{equation}
    \begin{equation}
        \beta_{\mathrm{PPN}}\equiv\frac{d\alpha_{\mathrm{PPN}}(\varphi_{*})}{d\varphi_{*}}=\alpha''\left(\varphi_{*}\right)=-\frac{\gamma}{\varphi_{*}^{2}}=-\frac{2\gamma}{\Phi^{2}}.
    \end{equation}
The post-Newtonian parameters\footnote{$\gamma^{\mathrm{PPN}}$ - measures the amount of curvature produced by a unit mass; $\beta_{\mathrm{PPN}}$ - quantifies the nonlinearity in the superposition law of gravity.} in terms of today's $\alpha_{\mathrm{PPN}}$ and $\beta_{\mathrm{PPN}}$ parameter values take the following form:
    \begin{equation}
        \begin{dcases}
            \gamma^{\mathrm{PPN}}-1\equiv-\frac{2\alpha_{\mathrm{PPN}0}^{2}}{1+\alpha_{\mathrm{PPN}0}^{2}}=-\frac{4\gamma^{2}}{2\gamma^{2}+\Phi^{2}_{0}}, \\
            \beta^{\mathrm{PPN}}-1\equiv\frac{\beta_{\mathrm{PPN}0}\cdot\alpha_{\mathrm{PPN}0}^{2}}{2\left(1+\alpha_{\mathrm{PPN}0}^{2}\right)^{2}}=-\frac{2\gamma^{3}}{\left(2\gamma^{2}+\Phi_{0}^{2}\right)^{2}}.
        \end{dcases}
    \end{equation}
In light of the fact that the experimentally obtained ranges of values for the aforementioned parameters are as follows \cite{Bertotti2003,Williams2004}:
    \begin{equation}
        \begin{dcases}
            \gamma^{\mathrm{PPN}}-1=\left(2.1\pm 2.3\right)\times 10^{-5}, \\
            \beta^{\mathrm{PPN}}-1=\left(1.2\pm 1.1\right)\times 10^{-4},
        \end{dcases}
\end{equation}
this allows an investigation to ascertain the compatibility or incompatibility of the models considered in this article with the available experimental data.

In consideration of the models under examination, the current PPN parameters values can be identified with the following numbers:
\begin{equation}
    \begin{dcases}
            \left|\gamma^{\mathrm{PPN}}-1\right|\simeq 2.68\times 10^{-7}, \\
            \left|\beta^{\mathrm{PPN}}-1\right|\simeq 3.66\times 10^{-14}.
    \end{dcases}
\end{equation}
The parameter values indicate the \textit{compatibility} of both models with the constraints imposed by the PPN formalism and BBN.

Furthermore, as the individual models evolve, the values of the aforementioned parameters lie within the following ranges:
\begin{itemize}
    \item Scalaron model:
        \begin{equation}
        \begin{dcases}
            2.68006\times 10^{-7}\leq\left|\gamma^{\mathrm{PPN}}_{\mathrm{sc}}-1\right|\leq 2.69154\times 10^{-7}, \\
            3.66465\times 10^{-14} \leq\left|\beta^{\mathrm{PPN}}_{\mathrm{sc}}-1\right|\leq 3.69611\times 10^{-14};
        \end{dcases}
        \end{equation}
    \item Phantom model:
        \begin{equation}
        \begin{dcases}
            2.64759\times 10^{-7}\leq\left|\gamma^{\mathrm{PPN}}_{\mathrm{ph}}-1\right|\leq 2.68006\times 10^{-7}, \\
            3.5764\times 10^{-14} \leq\left|\beta^{\mathrm{PPN}}_{\mathrm{ph}}-1\right|\leq 3.66465\times 10^{-14}.
        \end{dcases}
        \end{equation}
\end{itemize}
As one can observe, the consistency of the first PPN parameter is at the level of 2 orders of magnitude and the second at 10 orders of magnitude. Based on this evidence, it is reasonable to conclude that both models are consistent with the currently available observations.
\section{Discussion and perspectives}\label{Conclusions}

In this paper, we performed a detailed study of matter stress-energy non-conservation in ST FLRW universe and its relation with the chameleon mechanism. An explicit solution to this problem is proposed for any kind of dark fluid implemented by an arbitrary generating function $f$. Two toy models mimicking well LCDM one with the same predictions for the late universe (including supernovae data) and providing a realistic ratio between baryonic and dark matter were analyzed. Differences appear for the early universe ($a<10^{-1}$) which, particularly in the CMB epoch ($a\sim 10^{-3}$), requires studying of the cosmological perturbations. Preliminary results obtained here by numerical methods show that the scalaron case ($\epsilon=1$) suffers the same problems as LCDM: lack of internal inflationary mechanism. This can be cured by modification of self-interaction potential $\cv(\Phi)\neq \mathrm{const}$ and considering massive scalar field. In contrast, the phantom (ghost) case (ie. $\epsilon=-1$), as expected, offers (matter) bounce scenario instead of big bang.

In Section~\ref{Engineering of stress-energy}, we also pointed out the need to revise the description of the chameleon mechanism, in which in the general (non-barotropic) case we are not dealing only with a factor that multiplies the energy density of the material content of the Universe. With the help of the so-called generating function, we pointed out the general relations for generating different types of the so-called dark fluid, which treats dark matter and dark energy as manifestations of a single physical phenomenon. The aspect of these dark fluids (e.g., logotropic \cite{Benaoum2022} and Murnaghan \cite{Dunsby2024} fluids) regarding the chameleon mechanism seems worth considering in the future.
Reformulations of standard energy conditions for FLRW type cosmological models provided in the Subsection~\ref{Energy conditions and other applications} may provide an interesting scope for the study of energy conditions per se, as well as for uncommon yet physically interesting proposals of new forms of functions for energy densities in cosmological applications. These conditions are, in principle, a kind of assumptions and not constraints derived from fundamental principles in physics. For this reason, consideration of these conditions could lead to new interesting physical systems and discovering "new" laws of physics.

Of the two toy models we have obtained, the more promising seems to be the model belonging to the family of so-called Bounce Cosmology - EFSTG (Einstein Frame Scalar-Tensor Ghost) characterized by a matter (phantom (ghost) scalar field) bounce phase. It constitutes a particular example of an alternative hypothesis to the widely accepted model of cosmological inflation. However, the EFSTS model may be a realization of the description of the dark sector of the Universe known as the quintessence, which, when the more complicated scalar field potential is taken into account, may also exhibit interesting properties.

We have shown that in the case of our two toy models we cannot speak of their physical equivalence in the Einstein and Jordan frames. From our point of view, the Einstein frame (which does not interfere with the geometrical structure of gravity) is the physical one. The Jordan frame, on the other hand, in this case, is non-physical because it corresponds to a different physics. Furthermore, we have demonstrated the compatibility of both models with the constraints placed on the PPN parameters.

Scalar-tensor cosmological models in the Einstein frame, due to the chameleon mechanism, lead to non-conservation of the energy-momentum tensor \eqref{NMC  continuity}. From a physical point of view, this leads to a violation of the equivalence principle \cite{Hui2009}. At present, the WEP must be met at the level of $\eta\leq 10^{-15}$ (based on the estimation of the E{\"o}tv{\"o}s parameter from the MICROSCOPE mission \cite{Touboulfinal}). If the range of free parameters present in existing or future models violates this principle in such a limited manner, it will be possible to detect a new hypothetical fifth fundamental interaction related to the non-minimal (anomalous) coupling between the fundamental scalar field and matter. Notably, one such proposal could be models related to scalar dark matter, namely ultra-light (bosonic) DM \cite{Ferreira2021}. In such an approach the bosonic dark matter could manifest its existence \cite{Stadnik2023,Guo2023} through a recently identified problem with the anomalous magnetic dipole moment of the muon particle \cite{Muon}.

As it has been demonstrated in \cite{Cook2020}, the only known \textit{supersmoother} (classical, quantum, robust and rapid smoother \cite{Cook2020,Ijjas2018a}) in modern relativistic cosmology is the  phase of slow contraction (ekpyrotic phase \cite{Khoury2001}). When it comes to inflation, the inflaton field quantum fluctuations generate growing mode curvature fluctuations which, consequently, does not allow for the homogeneity of the Universe in the broadest sense. In order for the inflation mechanism to fulfill its primary purpose, we must deal with a strong narrowing of the possible values of the free parameters associated with the self-interaction scalar field potential and a narrow range of allowed inflaton initial velocities. As a result, we face the problem of \textit{fine-tuning} and \textit{initial conditions} \cite{Garfinkle2023}. In contrast, in models with a non-singular cosmological bounce, both issues do not arise. Instead of an initial singularity, we have an ekpyrotic (ultra-slow contraction) phase (geodesic completeness in mathematical sense), followed by a non-singular bounce, and then there is a release of some of the scalar field energy consumed in the reheating process - this aspect regarding the models introduced in our work should be further examined through the methods of perturbation theory (e.g. via quasi-static approximation in STT \cite{Pogosian2023}) and dynamical systems, especially regarding the formation of the large-scale structure of the Universe in terms of SFDM models \cite{Matos2000SFDM,Matos2000,Matos2001}.
    
Furthermore, such an evolution could generate a nearly scale-invariant spectrum of nearly gaussian density perturbations \cite{Finelli2002,Ijjas2018a} as in the inflationary scenario. This is one of the crucial observational criteria regarding modern cosmological models. In addition, from a thermodynamics perspective, the cyclic scenario of the Bounce Cosmology \cite{Ijjas2019} could avoid the well known Tolman entropy problem \cite{Tolman1934} plaguing earlier attempts to describe cyclic Universe scenario. It is also a challenge to try to introduce a non-singular (for $H=0$) description of entropy during a cosmological bounce (e.g., \cite{Odintsov2023Entropy}), which would allow an attempt to gain a deeper understanding of alternatives to cosmological inflation within the foundations of thermodynamics. These issues regarding the formalism proposed in our work should also be addressed in future research.

\section*{CRediT authorship contribution statement}
\textbf{Andrzej Borowiec:} Conceptualization, Formal Analysis, Investigation, Methodology, Validation, Writing – original draft, Writing – review \& editing.

\textbf{Marcin Postolak:} Conceptualization, Formal Analysis, Investigation, Methodology, Software, Validation, Visualization, Writing – original draft, Writing – review \& editing.

\section*{Declaration of competing interest}
The authors declare that they have no known competing financial interests or personal relationships that could have appeared to
influence the work reported in this paper.

\section*{Data availability}
No data was used for the research described in the article.

\section*{Acknowledgements}
This article is based upon work from COST Action CA21136 – “Addressing observational tensions in cosmology with systematics and fundamental physics (\href{https://cosmoversetensions.eu/}{CosmoVerse})”, supported by COST (European Cooperation in Science and Technology). AB is  supported by the project UMO-2022/45/B/ST2/01067 from the Polish National Science Center (NCN).

MP would like to express his sincerest appreciation to Roksana Szwarc for her very helpful and accurate remarks on numerical and symbolic calculations in Wolfram Mathematica, and to Alexander Kozak for inspiring discussions regarding scalar-tensor theories of gravity.

We would also like to express our sincere appreciation to Orlando Luongo for his valuable comments regarding dark fluid models and his interest in our article.

Furthermore, the authors express their gratitude to the Reviewer of the article for valuable suggestions and commentary.



\biboptions{sort&compress}
\bibliographystyle{elsarticle-num}
\bibliography{main}





\end{document}